**FP-DeErr: Application-Oriented Error Decomposition for Foundation Potentials**

Kiyan Amirian, Ramanuja Srinivasan Saravanan, Felix Adams, Charles E Schwarz, Yifei Mo*

Department of Materials Science and Engineering, University of Maryland, College Park, MD 20742, USA

* Email: yfmo@umd.edu

**Abstract.** Foundation potentials (FPs) have emerged as a new basis for atomistic modeling. While their evaluation using average energy and force errors often indicates near-DFT accuracy, their performance in practical computational studies remains inconsistent. While recent benchmarks evaluate FPs on downstream computational tasks, the underlying errors that determine task success or failure are not always clear. Here, we develop FP-DeErr, an application-oriented error-decomposition framework that resolves FP errors according to the physically meaningful quantities, configurations, and computational stages governing specific tasks. Rather than averaging errors over an entire dataset, FP-DeErr uses decomposed error metrics devised for physically meaningful quantities, focusing on the configurations where errors arise and matter most. We demonstrate FP-DeErr by evaluating multiple state-of-the-art FPs on three fundamental tasks: force prediction for atomistic simulations, energy ranking for substitutional and vacancy orderings, and ion/vacancy migration. These error-decomposition metrics resolve force errors among highly accurate, large-error, and far-from-equilibrium atoms; relative-energy errors among competing orderings, phases, and compositions; and ion migration errors among endpoints and along-path errors. By identifying where FP errors arise, this error decomposition provides targeted guidance for FP development. FP-DeErr also provides an open benchmark, evaluation code, and a public leaderboard for rigorous FP assessment.

## 1. Introduction

Density functional theory (DFT) has been the standard approach for computational materials studies across diverse material systems, but its high computational cost limits simulations of large systems and long timescales. Machine-learning interatomic potentials (MLIPs) provide orders-of-magnitude acceleration by learning from DFT data,[1] and recent graph neural network-based universal MLIPs, often referred to as foundation potentials (FPs), extend this approach from system-specific potentials to broad chemical and structural coverage through training on large materials datasets such as the Materials Project (MP).[2,3] Early FP models of M3GNet[2], CHGNet,[4] and MACE,[5] are largely trained on DFT structure-relaxation data, which are heavily biased towards near-equilibrium configurations. However, many computational tasks using FPs involve far-from-equilibrium configurations encountered in defects, diffusion, and finite-temperature dynamics.[6–9] Newer datasets, including OMat24[10] and MatPES[11], expand the sampling of the potential energy surface (PES), and enable newer FP models, including UMA[12], MatterSim,[13] Orb,[14] DPA4,[15] GRACE,[16] and MatPES-trained FPs.[11] Owing to their broad applicability across solid materials, FPs are increasingly serving as DFT surrogates or preprocessing methods in computational materials workflows, including structural relaxation, molecular dynamics (MD) simulation, phase-diagram construction, defect calculations, and nudged elastic band (NEB) simulations.[8,9,17]

FPs are still commonly evaluated and ranked using average error metrics, such as energy and force mean absolute error (MAE) or root-mean-square error (RMSE), typically computed on structures that are in-distribution with the training data.[2,4,5] Although models with low MAE or RMSE are often regarded as having reached near-DFT accuracy, average error metrics alone do not indicate whether an FP is reliable for the downstream tasks for which it is used.[6,18–22] This is caused by three issues. (1) Average energy and force errors evaluate prediction accuracy over a dataset rather than the physical quantities that directly determine the outcome of a computational task. (2) Averaging errors over the full population of atoms or structures in the dataset can mask errors concentrated in a small, physically consequential subset of atoms or structures, an effect we refer to as the dilution effect. (3) Evaluation datasets often resemble the training distribution and are dominated by near-equilibrium structures,[2,4,5,11] so far-from-equilibrium configurations and near-degenerate competing phases or orderings are often underrepresented despite being critical to many practical computational tasks. For example, in system-specific MLIPs, Liu et al. showed that low average energy error does not necessarily imply accurate ranking of competing elemental orderings,[23] and that MLIPs with low average error can still perform poorly on rare, physically consequential events, such as diffusion.[7] For some FPs, systematic underestimation of migration barriers[19,24,25] and deviations in the local atomistic configurations along the migration pathways have been reported.[19,25] Thus, to address all these three issues, MLIPs, including FPs, should be evaluated on the physical quantities and configurations that govern performance in specific computational tasks.

Recent benchmarks have broadened FP evaluation beyond average errors to performance in practical computational tasks.[6,18,24–32] These evaluations span thermodynamic stability, structural relaxation, mechanical properties, defect and surface energetics, and phonons,[18,26–31] as well as finite-temperature structural and dynamical behavior.[6,32] These studies demonstrate the importance of evaluating FPs on physical properties and computational tasks. However, while an aggregate task-level score indicates whether an FP succeeds or fails, it may not reveal which underlying prediction causes the failure or where the error arises. Aggregate task-level scores can also be subject to the dilution effect when they average over configurations that differ substantially in physical consequence or prediction difficulty. Therefore, a further level of error decomposition is needed to resolve task-level performance according to the physical quantities, configurations, and computational stages in which consequential errors arise.

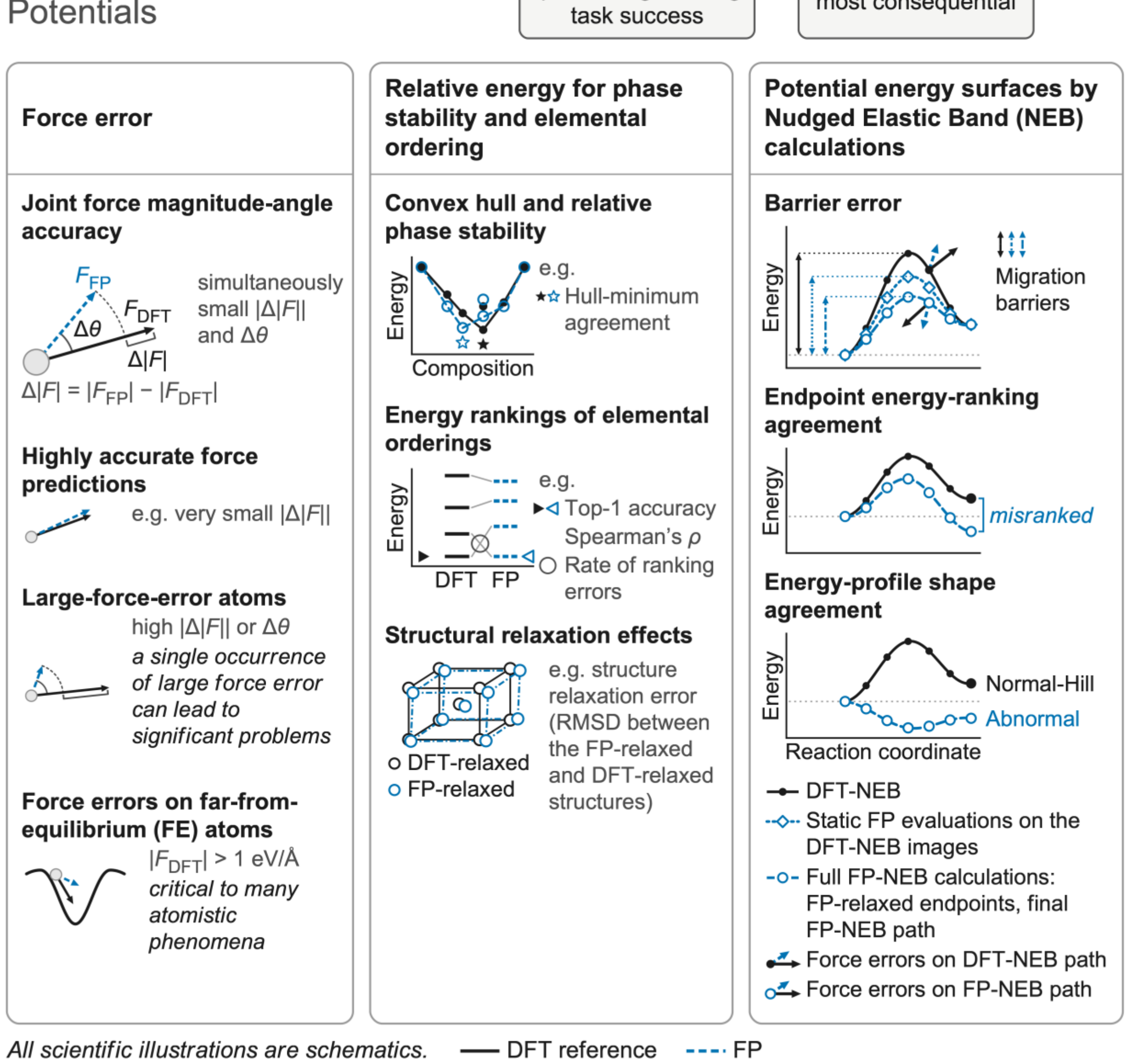

**Fig. 1: The FP-DeErr application-oriented error-decomposition framework.**

In this work, we develop FP-DeErr, an application-oriented error-decomposition framework (Fig. 1) for evaluating FP performance by decomposing errors according to the physical quantities, configurations, and computational stages that govern the outcome of a computational task. In this error-decomposition framework, the errors are evaluated using (1) decomposed error metrics specifically devised to quantify (2) the physically meaningful quantities governing task success, with a focus on (3) the configurations where errors are most consequential, rather than averaging errors over the full dataset or computational workflow. This FP-DeErr framework (Fig. 1) is demonstrated on three representative computational tasks: force prediction for atomistic simulations (Section 2.1), relative-energy prediction for phase stability and elemental ordering (Section 2.2), and ion migration by NEB calculations (Section 2.3). In Section 2.1, force-prediction errors are decomposed according to prediction accuracy and physically consequential atom subsets, including highly accurate, large-force-error, and far-from-equilibrium atoms. In Section 2.2, errors in thermodynamic-stability prediction are decomposed according to relative-energy comparisons among competing orderings, phases, and compositions, and further separated into intrinsic energy-ranking and structural-relaxation contributions. In Section 2.3, errors in FP-NEB calculations are decomposed according to the underlying PES and the stages of the FP-NEB workflow, including endpoint relaxation and along-path forces. Together, these analyses demonstrate how application-oriented error decomposition can diagnose FP failures that are not apparent from average errors or aggregate task-level performance.

## 2. Results

### 2.1. Force error

Currently, MLIP and FP force errors are commonly evaluated using MAE/RMSE over test datasets that are often in-distribution with the training data. For force predictions, we decompose errors according to force magnitude and direction, prediction-accuracy regimes (highly accurate and large-force-error predictions), and physically consequential atom subsets (far-from-equilibrium atoms), using the metrics summarized in Table 1. We evaluate these decomposed force errors using metrics (Table 1) for eleven FP models (Methods) on the MatPES dataset (and on the OMat24 rattled-1000 dataset in Supplementary Table 2). The analyses are performed on the training dataset for some models but are out-of-distribution for others, nonetheless, revealing key error sources and failure modes beyond average MAE/RMSE.

*2.1.1. Non-negligible force errors despite low average MAE*

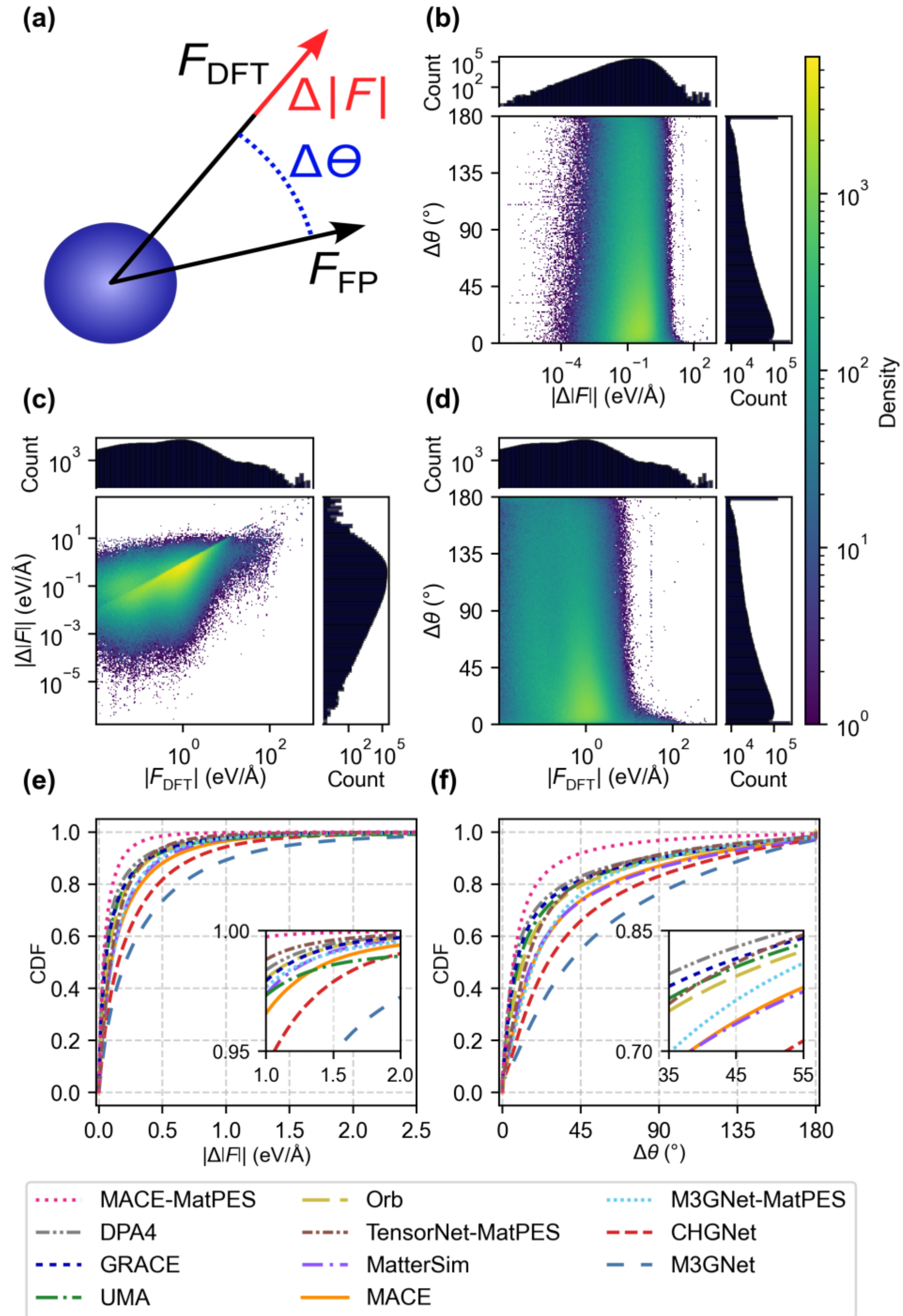

**Fig. 2: Force error analysis of FPs.** (a) Schematic illustrating the force-magnitude error ($\Delta|F| = |F_{\mathrm{FP}}| - |F_{\mathrm{DFT}}|$) and force-angle error ($\Delta\theta$) of FP-predicted force $F_{\mathrm{FP}}$ relative to the DFT force $F_{\mathrm{DFT}}$. (b–d) Density heatmaps for M3GNet with marginal histograms: (b) $|\Delta|F||$ versus $\Delta\theta$; (c, d) $|\Delta|F||$ and $\Delta\theta$ as functions of $|F_{\mathrm{DFT}}|$, respectively. Cumulative distribution functions (CDFs) of (e) $|\Delta|F||$ and (f) $\Delta\theta$ for all FP models, with insets enlarging the large-force-error region.

**Table 1:** Error-decomposition metrics for force prediction (details are provided in the Methods).

| Name | Metrics |
|---|---|
| Average error | Force-magnitude error $\Delta\|F\|$ and force-angle error $\Delta\theta$, reported as MAE/RMSE over all atoms or a selected subset. |
| Cumulative distribution functions (CDFs) of force errors | CDFs of $\|\Delta\|F\|\|$, $\Delta\theta$, and norm of the force-vector error $e_{\mathrm{vec}}$, over all atoms or a selected subset. |
| Highly accurate force predictions (small-force-error atoms) | Fraction of atoms with very small values of $\|\Delta\|F\|\|$ and $\Delta\theta$ below threshold (e.g. $\|\Delta\|F\|\| < 0.01$ eV/Å). |
| Joint force magnitude-angle accuracy | Fraction of atoms with simultaneously small values of $\|\Delta\|F\|\|$ and $\Delta\theta$ (e.g. $\|\Delta\|F\|\| < 0.01$ eV/Å and $\Delta\theta < 1°$ or 20°). |
| Large-force-error atoms | Fraction of atoms with high values of $\|\Delta\|F\|\|$ and $\Delta\theta$ (e.g. $\|\Delta\|F\|\| > 0.5$ eV/Å). |
| Force errors on far-from-equilibrium (FE) atoms | MAE/RMSE evaluated for $\Delta\|F\|$, $\Delta\theta$ over the FE atoms selected as $\|F_{\mathrm{DFT}}\| > 1$ eV/Å. Fraction of FE atoms with relative force-magnitude error $r_F$ below increasing thresholds. |

Underlying the overall average error, the distributions of force-magnitude error $|\Delta|F||$ versus force-angle error $\Delta\theta$ (Fig. 2b-d) show a high-density region around $|\Delta|F|| \lesssim 0.1$ eV/Å, indicating reasonably accurate force-magnitude predictions for a high fraction of atoms in the evaluation dataset, consistent with the low MAE/RMSE over all atoms. The CDFs of $|\Delta|F||$ (Fig. 2e) and $\Delta\theta$ (Fig. 2f) provide a quantitative view of the force-error distributions, as previously used in Ref.[7]

For all FP models, 73-98% of atoms show force-magnitude errors of $|\Delta|F|| < 0.5\ \mathrm{eV/Å}$ (Fig. 2e, Fig. 3). Beyond force magnitude, force-angle accuracy is also important[33]. For a relatively loose criterion of $\Delta\theta$ < 20°, M3GNet achieves 31.0% while UMA, GRACE and DPA4 achieve 66–71% and MACE-MatPES reaches 80.3% (Supplementary Fig. 4). There is a notable fraction of atoms showing large force-magnitude errors $|\Delta|F||$ or large force-angle errors $\Delta\theta$ (Fig. 2b-d, Fig. 4 and Supplementary Fig. 4), and only a small fraction of atoms satisfy the joint force magnitude-angle accuracy criterion (Fig. 3).

**Table 2:** Results of key metrics for each FP model (Table 1 and Methods). All metrics are evaluated for atoms with $|F_{\mathrm{DFT}}| > 0.01$ eV/Å, or for the FE-atoms with $|F_{\mathrm{DFT}}| > 1$ eV/Å. Metric-specific thresholds are given in the column headers. Results for the OMat24 rattled-1000 dataset are provided in Supplementary Table 2.

| **FP models** | **$\Delta\|F\|$ MAE/ RMSE (eV/Å)** | **$\Delta\theta$ MAE/RMSE (deg)** | **$\Delta\|F\|$ MAE/RMSE (eV/Å), $\|\Delta\|F\|\|$ < 1 eV/Å** | **Frac. $\|\Delta\|F\|\|$ < 0.01 eV/Å (%)** | **Frac. $\|\Delta\|F\|\|$ < 0.01 eV/Å & $\Delta\theta$ < 1°/20° (%)** | **Frac. $\|\Delta\|F\|\|$ > 1 eV/Å (%)** | **$\Delta\|F\|$ MAE/RMSE (eV/Å), on FE atoms** | **$\Delta\theta$ MAE/RMSE (deg), on FE atoms** |
|---|---|---|---|---|---|---|---|---|
| MACE | 0.23 / 1.93 | 37 / 58 | 0.17 / 0.27 | 8.3 | 1.1 / 4.5 | 3.44 | 0.49 / 3.65 | 18 / 29 |
| CHGNet | 0.31 / 2.19 | 45 / 66 | 0.22 / 0.32 | 5.6 | 0.66 / 2.2 | 5.71 | 0.70 / 4.14 | 24 / 36 |
| M3GNet | 0.44 / 1.49 | 56 / 76 | 0.26 / 0.36 | 4.6 | 0.54 / 1.5 | 10.79 | 0.98 / 2.76 | 40 / 56 |
| MatterSim | 0.21 / 1.29 | 37 / 59 | 0.17 / 0.26 | 7.9 | 1.00 / 4.5 | 2.65 | 0.40 / 2.41 | 16 / 28 |
| UMA | 0.56 / 7.93 | 28 / 51 | 0.12 / 0.21 | 15.1 | 2.8 / 11 | 2.69 | 1.13 / 11.89 | 12 / 25 |
| Orb | 0.17 / 1.12 | 30 / 52 | 0.14 / 0.22 | 12.6 | 1.4 / 8.9 | 2.00 | 0.31 / 2.09 | 12 / 24 |
| DPA4 | 0.15 / 2.16 | 25 / 48 | 0.11 / 0.20 | 19.7 | 4.7 / 16 | 1.64 | 0.27 / 3.71 | 10 / 22 |
| GRACE | 0.16 / 1.49 | 27 / 50 | 0.12 / 0.21 | 17.8 | 3.7 / 14 | 2.08 | 0.30 / 2.81 | 12 / 25 |
| M3GNet-MatPES | 0.21 / 1.24 | 34 / 53 | 0.17 / 0.25 | 7.1 | 0.92 / 3.7 | 2.35 | 0.43 / 2.34 | 18 / 27 |
| TensorNet-MatPES | 0.17 / 1.39 | 29 / 49 | 0.15 / 0.22 | 7.7 | 0.98 / 4.6 | 1.20 | 0.33 / 2.63 | 12 / 18 |
| MACE-MatPES | 0.08 / 1.10 | 16 / 32 | 0.07 / 0.13 | 19.0 | 3.9 / 15 | 0.25 | 0.16 / 2.09 | 6 / 10 |

### *2.1.2. Highly accurate force predictions: small-force-error atoms*

Highly accurate force predictions are desired because a strict force convergence criterion of 0.001–0.01 eV/Å is often used to relax structures for subsequent computations such as defect energies, electronic structure, and NEB. Since relaxation is directly guided by the predicted forces, force errors comparable to these convergence thresholds can drive the relaxation toward a different minimum or cause it to stop before reaching the correct DFT equilibrium structure. Yet, for most FP models, only a small fraction (<10%) of atoms meet the strict accuracy criterion of $|\Delta|F|| < 0.01$ eV/Å with 4.6% for M3GNet and 15% for UMA across the dataset (Fig. 3). DPA4 gives the highest fraction among the tested models, reaching 19.7%, followed by MACE-MatPES at 19%.

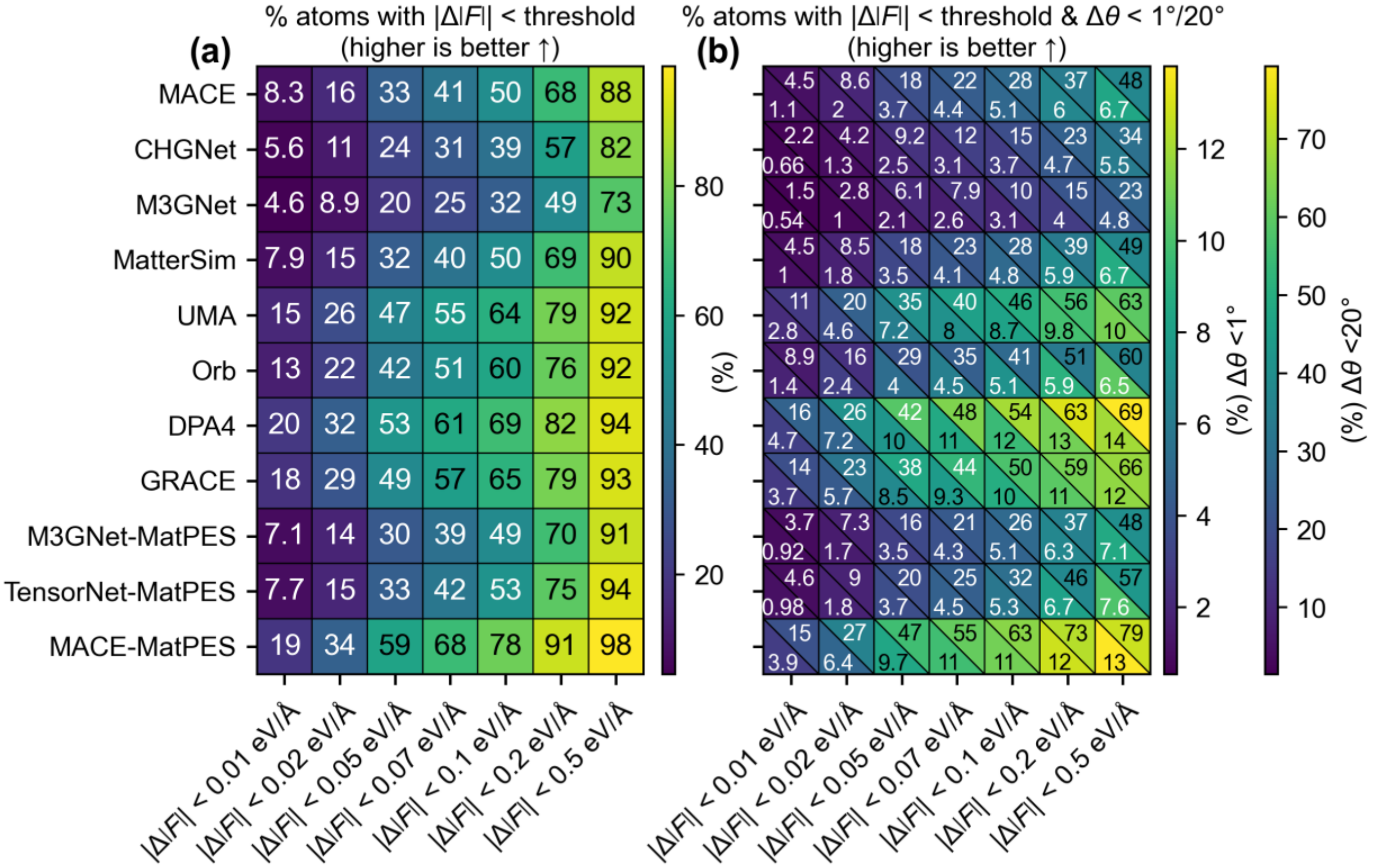

**Fig. 3: Highly accurate force predictions and joint force magnitude-angle accuracy.** (a) The fraction of atoms with small force-magnitude errors ($|\Delta|F||$) below the thresholds. (b) The fraction of atoms with simultaneously small force-magnitude error ($|\Delta|F|| <$ threshold) and force-angle error ($\Delta\theta < 1°$, lower triangle; $\Delta\theta < 20°$, upper triangle). Analyses performed on the OMat24 rattled-1000 dataset and for the force-vector error $e_{\text{vec}}$ are provided in Supplementary Table 2 and Supplementary Fig. 1, respectively.

Approximately 1% or fewer of atoms meet the stricter joint force magnitude-angle accuracy criterion of $|\Delta|F|| < 0.01$ eV/Å and $\Delta\theta < 1°$ for most FP models (Fig. 3). Even for the better-performing models, this fraction remains small, reaching only 3.7% for GRACE and 4.7% for DPA4. Under the less strict criteria of $|\Delta|F|| < 0.1$ eV/Å and $\Delta\theta < 20°$, M3GNet and CHGNet reach only 10–15%, while UMA, GRACE and DPA4 achieve 46–54% and MACE-MatPES reaches 63% (Fig. 3). This low fraction demonstrates the difficulty of simultaneously predicting force magnitude and direction with high accuracy.

*2.1.3. Large-force-error atoms*

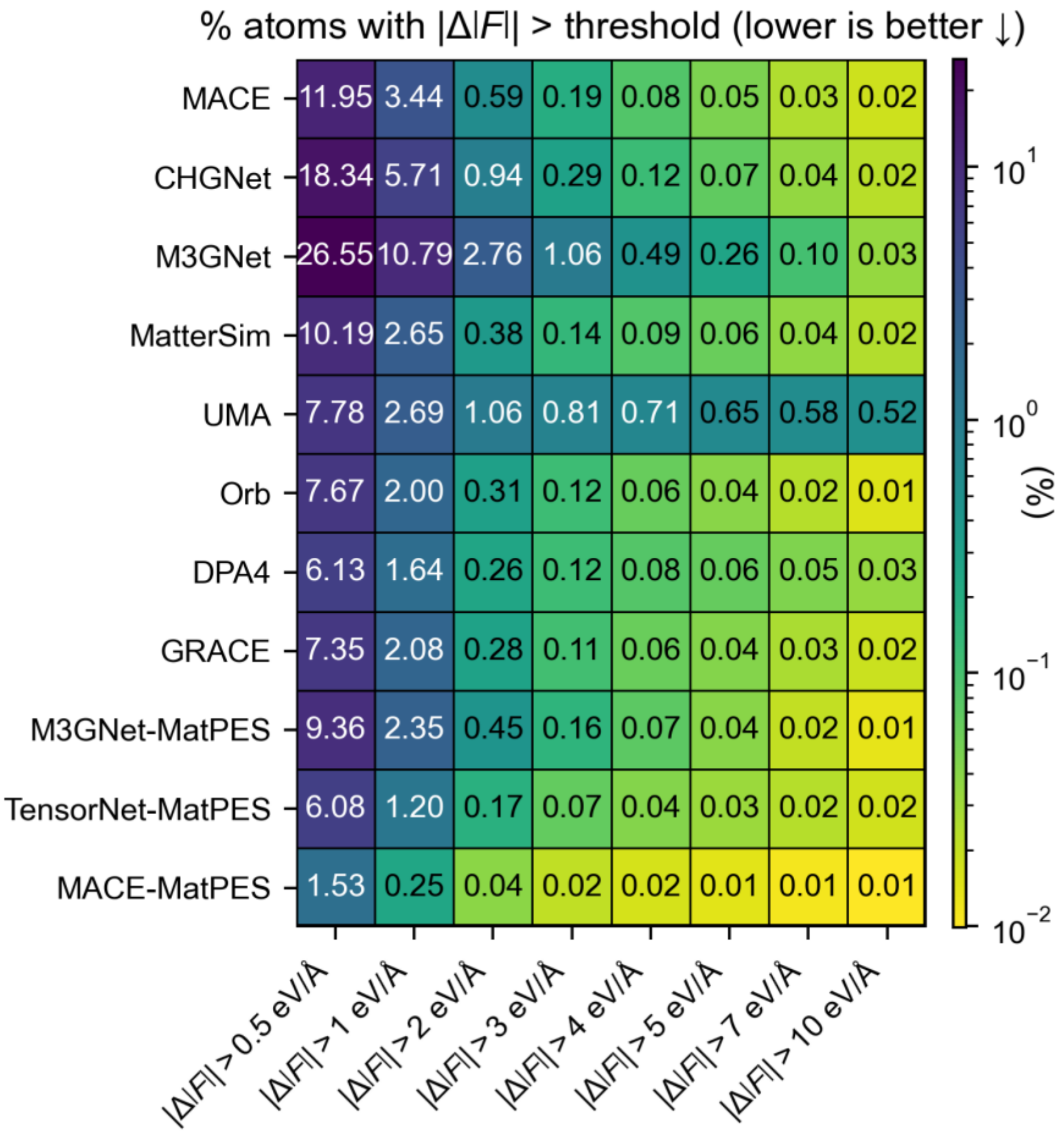

**Fig. 4: Large-force-error atoms.** Fraction of atoms with force-magnitude errors $|\Delta|F||$ exceeding different large-error thresholds.

Large force errors are highly undesired in atomistic simulations using FPs, because even a single occurrence of large force error can lead to significant problems, such as erroneous atomic dynamics in MD simulations or failure of convergence in NEB calculations.[20,22,32] A non-negligible fraction of atoms exhibit such large force-magnitude errors (Fig. 2e, Fig. 4) and force-angle errors (Fig. 2f, Supplementary Fig. 4). At the large force-magnitude error threshold of 0.5 eV/Å or 1 eV/Å, the corresponding fractions are 26.55% and 10.79% for M3GNet, 11.95% and 3.44% for MACE, and 1.53% and 0.25% for MACE-MatPES (Fig. 4). Even higher force-magnitude errors of $|\Delta|F||$ >10 eV/Å are observed for some atoms across all FP models (Fig. 4). Large force-angle errors are also notable: for $\Delta\theta > 90°$, M3GNet reaches 23.5% while MACE-MatPES is lowest at 3.2% (Supplementary Fig. 4). The fraction of large-force-error atoms is another key error metric for FPs.

It is worth noting that, while UMA has one of the highest fractions of small force-magnitude errors among the FPs (Fig. 2e, f, and Fig. 3), it also has the highest fraction of large force-magnitude errors (e.g. 0.65% at $|\Delta|F|| > 5$ eV/Å) (Fig. 4). The larger fraction of large force-magnitude errors is also reflected in the increased difference between MAE and RMSE in UMA (Fig. 5), as the higher fraction of these large-error atoms contributes significantly more to RMSE.

### *2.1.4. Far-from-equilibrium (FE) atoms*

Many computational tasks involve far-from-equilibrium (FE) atoms, such as migrating ions, defect and disordered structures, and large thermal displacements in high-temperature MD simulations. Since near-equilibrium atoms experience a relatively small force magnitude, we distinguish FE atoms simply using a DFT force magnitude cutoff of $|F_{\mathrm{DFT}}| > 1$ eV/Å. Approximately 26.8% of atoms evaluated in the MatPES dataset (Methods) are classified as FE atoms (Fig. 5). For comparison, that fraction is 6.1% for the MPTrj dataset (Supplementary Table 3), which mostly focuses on near-equilibrium relaxation of atomic structures.[11,34]

For those FP models trained on the structure-relaxation data, which primarily includes near-equilibrium structures, the force error is significantly larger for FE atoms ($|F_{\mathrm{DFT}}| > 1$ eV/Å) than the MAE/RMSE over all atoms (Fig. 5). The FP models trained on the MatPES dataset, such as M3GNet-MatPES and TensorNet-MatPES, perform considerably better for FE atoms with an MAE of 0.43 and 0.33 eV/Å, respectively, compared to 0.49–0.98 eV/Å for those models trained on the DFT structure-relaxation dataset (Fig. 5). MACE, although trained on MPTrj and sAlex datasets, performs comparably to the MatPES-trained models. MatterSim, Orb, DPA4 and GRACE achieve FE force-magnitude MAEs of 0.27–0.40 eV/Å, as their training data contains off-equilibrium configurations. MACE-MatPES shows the strongest FE performance among these models, reaching the lowest MAE of 0.16 eV/Å.

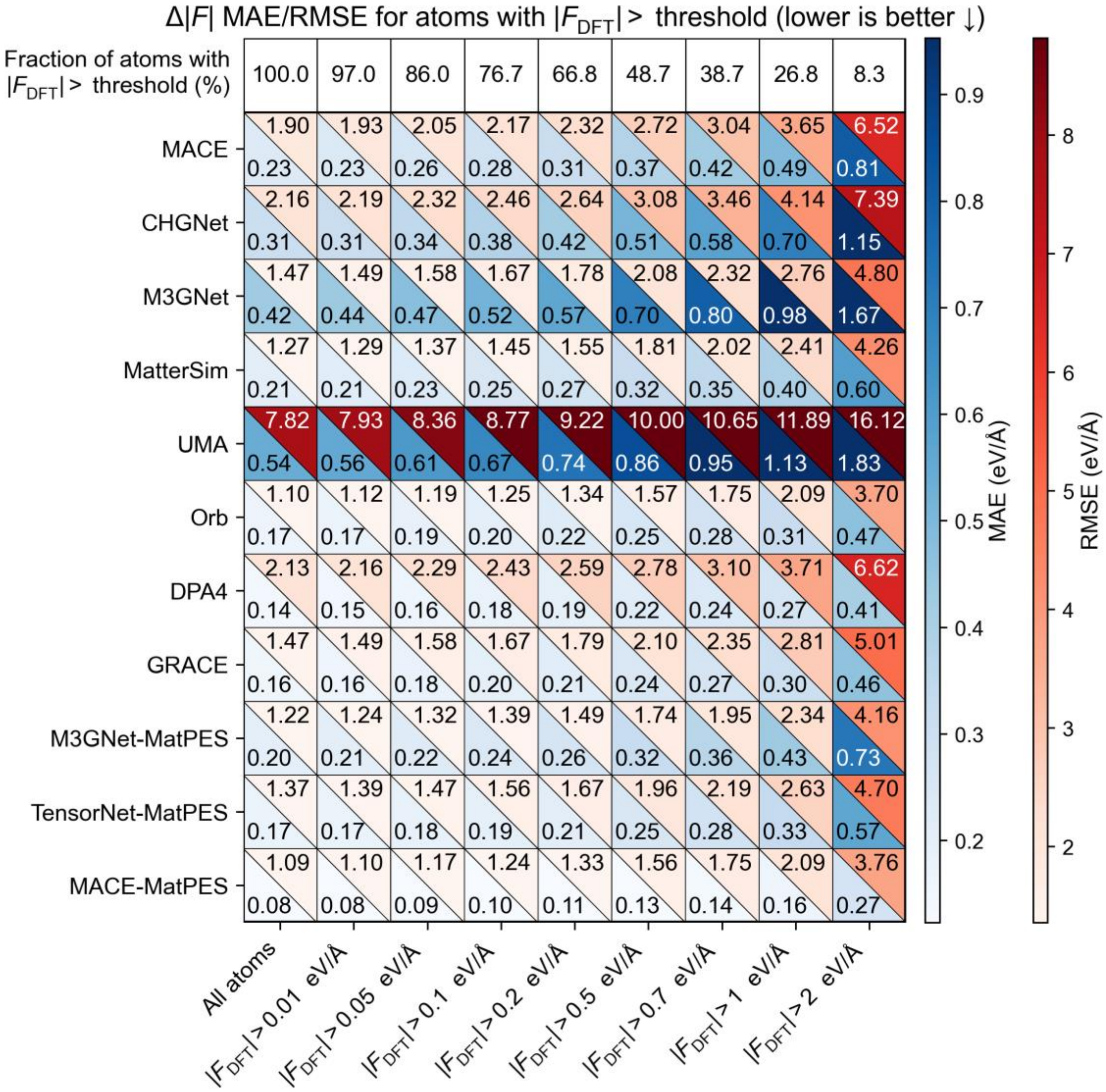

**Fig. 5: Average force-magnitude error.** $\Delta|F|$ MAE (blue, lower-left triangles) and RMSE (red, upper-right triangles) for subsets of atoms with $|F_{\mathrm{DFT}}|$ greater than increasing thresholds. Increasing the threshold excludes near-equilibrium and focuses the analysis on far-from-equilibrium (FE) atoms, with the top row showing the fraction of atoms at each threshold.

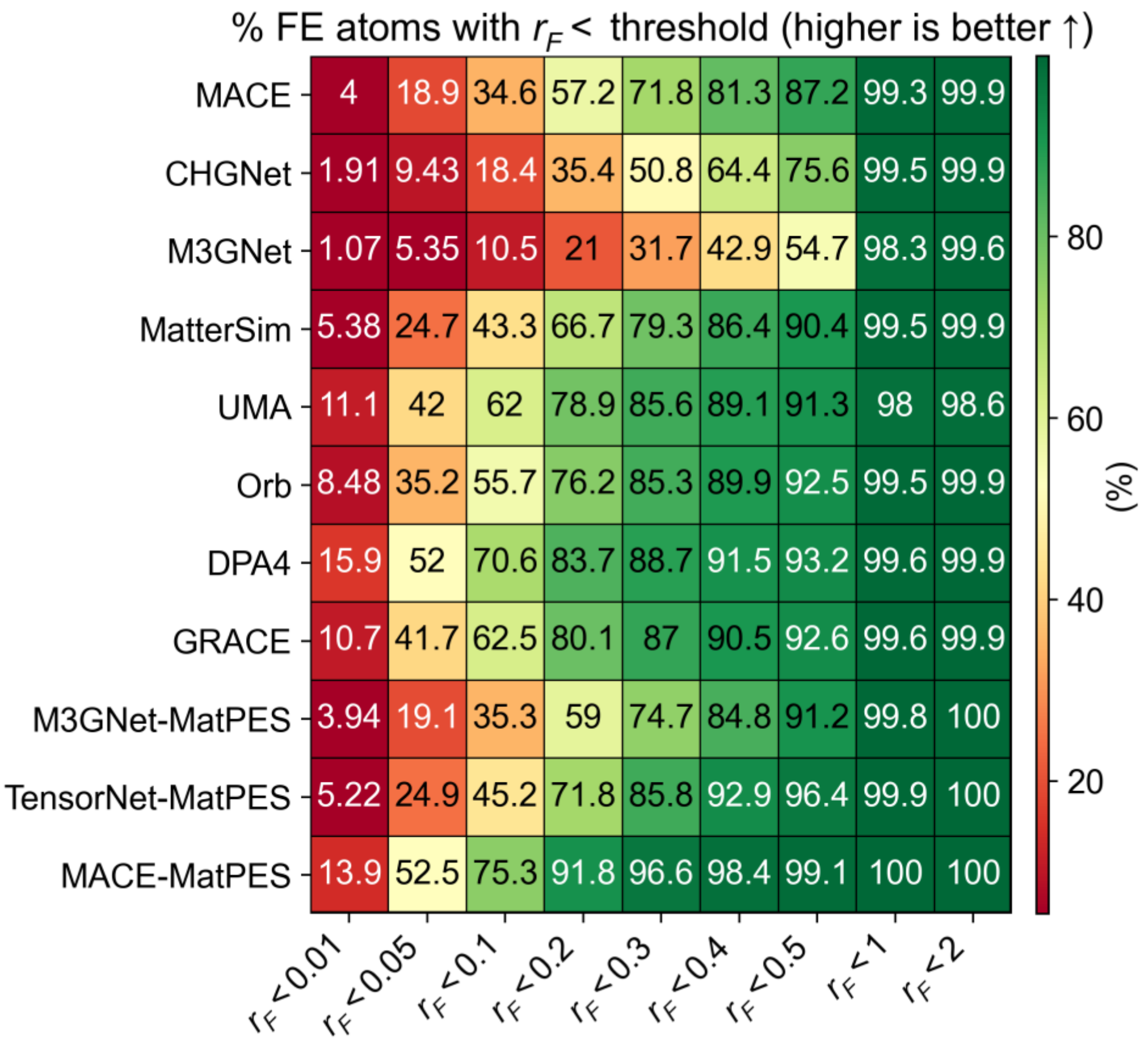

**Fig. 6: Force errors on far-from-equilibrium (FE) atoms.** Fraction of FE atoms ($|F_{\mathrm{DFT}}| > 1$ eV/Å) with relative force-magnitude error $r_F$ below the thresholds (Table 1 and Methods).

The stronger FE performance of MatPES-trained models is expected, given the broader sampling of far-from-equilibrium configurations in MatPES,[11] and these MatPES-trained models are evaluated on their training data. On OMat24 rattled-1000 dataset, the models trained or pre-trained on OMat24 (UMA, DPA4, Orb and GRACE) give the lowest FE-atom MAEs of 0.10–0.20 eV/Å, while the MatPES-trained models rise to 0.51–0.70 eV/Å (Supplementary Table 2). These results demonstrate that training on non-equilibrium structures can improve performance on FE atoms. However, UMA, which was trained on the OMat24 dataset including many non-equilibrium structures,[10] shows the highest FE-atom force-magnitude MAE of 1.13 eV/Å on the MatPES dataset, and a notably high fraction of large relative error of $r_F > 2$ (1.4%) among FE atoms, compared to 0–0.4% for other FPs (Fig. 6).

Atoms not in the FE-atom subset show generally similar errors. For the remaining 73.2% of atoms with $|F_{\mathrm{DFT}}| \leq 1$ eV/Å (i.e. the non-FE atoms) in the MatPES dataset (Fig. 5), only a small fraction of atoms are highly accurate (Supplementary Figs. 8 and 9). For UMA and M3GNet, 43.4% and 30.3% of their large force-magnitude errors with $|\Delta|F|| >$ 0.5 eV/Å originate from non-FE atoms (Supplementary Fig. 12a).

Given the importance of the FE atoms for atomistic simulations of many physical processes, force errors on FE atoms are a particularly important metric. To evaluate them, the analysis should be restricted to FE atoms, which may be a subset screened from the test dataset; otherwise, these errors are diluted by the dominant near-equilibrium atoms.

*2.1.5. Model comparison*

The models trained directly on the MatPES dataset perform strongly overall, although it should be noted that they are evaluated for errors on their training data. Even so, our error metrics reveal substantial errors in challenging cases that remain difficult even within the training data, providing targets for future model improvement.

MACE performs competitively on the out-of-distribution MatPES and OMat24 rattled-1000 datasets. Its performance across several key metrics, such as highly accurate force predictions, joint force magnitude-angle accuracy, and large-force-error atoms, is comparable to that of MatPES-trained models. Among our tested models, MACE-MatPES performs strongly across all force-error metrics on the MatPES dataset, although this performance does not transfer to the OMat24 rattled-1000 subset (Supplementary Table 2), despite OMat24 being included in its pretraining.

UMA also performs strongly on several metrics, with 15% highly accurate forces ($|\Delta|F||$< 0.01 eV/Å) and about 11% joint force magnitude-angle accuracy under a loose angle criterion ($\Delta\theta < 20°$), ranking behind DPA4, MACE-MatPES and GRACE in this metric on the MatPES dataset. However, UMA simultaneously exhibits one of the highest fractions of large-force-error atoms, reaching 0.65% at $|\Delta|F|| >$ 5 eV/Å (Fig. 4). The large gap between MAE and RMSE also reflects the larger fraction of large-force-error atoms, which contribute more to RMSE. Given UMA's strong performance for highly accurate and typical force predictions, its poorer performance on large-force-error cases warrants further investigation. Notably, these large force errors are not observed on the OMat24 rattled-1000 (Supplementary Table 2), suggesting that UMA's performance in large-force-error atoms on OMat24 does not fully transfer to the MatPES dataset. By contrast, M3GNet exhibits a lower fraction of large-force-error atoms than UMA at high thresholds (e.g. $|\Delta|F|| > 4$ eV/Å, Fig. 4), despite performing worse on other metrics such as highly accurate force predictions. In practical applications of FPs, having a low occurrence of large-force-error atoms is highly desired, as such predictions can lead to catastrophic failure in computational tasks or MD simulations.[20,22,32]

Through this error decomposition, the force prediction is resolved into distinct various force-prediction failure modes that are obscured by MAE/RMSE over all atoms (Table 2). In particular, large-force-error and FE-atom errors are physically consequential but can be diluted in the average error by the dominant population of near-equilibrium, low-error atoms (Fig. 5). Similarly, highly accurate force predictions are critical for many computational tasks but contribute little to the overall average error and can therefore be overlooked. Together with CDFs of force-magnitude and force-angle errors (Fig. 2e, f), these metrics (Table 1) identify which aspects of force prediction require improvement even when models exhibit similar average errors.

### 2.2. Relative energy for phase stability and elemental ordering

FPs are increasingly deployed in practical computational materials studies involving competing phases, substitutional and vacancy orderings, and defect structures beyond ideal bulk crystals.[9,17] Two representative applications are (i) constructing convex hulls,[28,35] and (ii) identifying the thermodynamically preferred substitutional, vacancy, and defect orderings.[23] In these applications, a key physically meaningful quantity is the relative stability of competing phases, compositions, and elemental orderings, whose energy differences may be only a few meV/atom.[23,36] Average energy error over all structures does not directly evaluate whether these relative stabilities are preserved.[23,26] Here, our error-decomposition metrics (Tables 3 and 5) evaluate thermodynamic-stability prediction at three levels: relative energies across competing phases and compositions, energy rankings among elemental orderings at fixed phase and composition, and the effects of structural relaxation.

As a realistic benchmark for these tasks, we use a dataset of ternary chalcogenide phase-change-material (PCM) systems from the first-principles thermodynamic study of Adams et al.[36] The dataset spans the composition space of Si, Ge, Sn, Sb, Bi, Ga, In and Ti selenides and tellurides, with structures generated by interpolating ternary mixture compositions along binary–binary chalcogenide tie lines (Methods). Each tie-line system contains multiple phases and their mixing compositions, each with multiple elemental orderings from isovalent cation or anion substitutions or aliovalent cation substitutions with charge compensation by cation or anion vacancies. This structural diversity in this dataset spans competing polymorphs, tie-line compositions, and elemental/vacancy orderings, and thus provides an application-oriented, physically motivated test for FPs. In representative Ge–Sb–Te systems, the DFT ground-state hexagonal phase and the metastable rock-salt phase differ by less than 10 meV/atom.[36] Such small energy differences make phase-stability and ordering predictions particularly demanding, because even modest FP energy errors can change the predicted ground state or energy ranking of elemental orderings.

Using this dataset, we decompose FP errors in thermodynamic-stability prediction into three parts. In Section 2.2.1, we test the accuracy of convex-hull construction and phase-stability prediction by comparing FP-predicted and DFT reference convex hulls across competing phases and compositions. In Section 2.2.2, we examine the relative-energy ranking of multiple elemental orderings within the same phase at fixed composition, using ranking metrics introduced by Liu et al.[23] together with complementary metrics. In all analyses, FPs are evaluated under two protocols: full FP relaxation and static FP evaluations. In full FP relaxation, each structure is relaxed with the FP, and the relaxed energy is used for convex-hull construction and ordering analysis, reflecting how FPs are used in practice. In static FP evaluations, the FP energy is computed directly on the DFT-relaxed structures without further relaxation (Methods). In Section 2.2.3, we compare full FP relaxation with static FP evaluations to separate intrinsic energy-ranking errors from those introduced by FP structural relaxation.

*2.2.1. Convex hull and relative phase stability*

Convex-hull construction provides a direct test of whether FPs can reproduce phase stability across competing phases and compositions. We first reduce the dataset to the lowest-DFT-energy ordering for each phase and composition forming the convex-hull subset (Methods). The FP-predicted energies of these structures are then used to construct FP convex hulls, which are compared directly with the DFT convex hulls as illustrated for the $GeSe_2$–$SiSe_2$ tie-line system in Fig. 7. The goal of this test is to evaluate whether the FP preserves the relative phase stabilities that determine ground states and convex-hull minima.

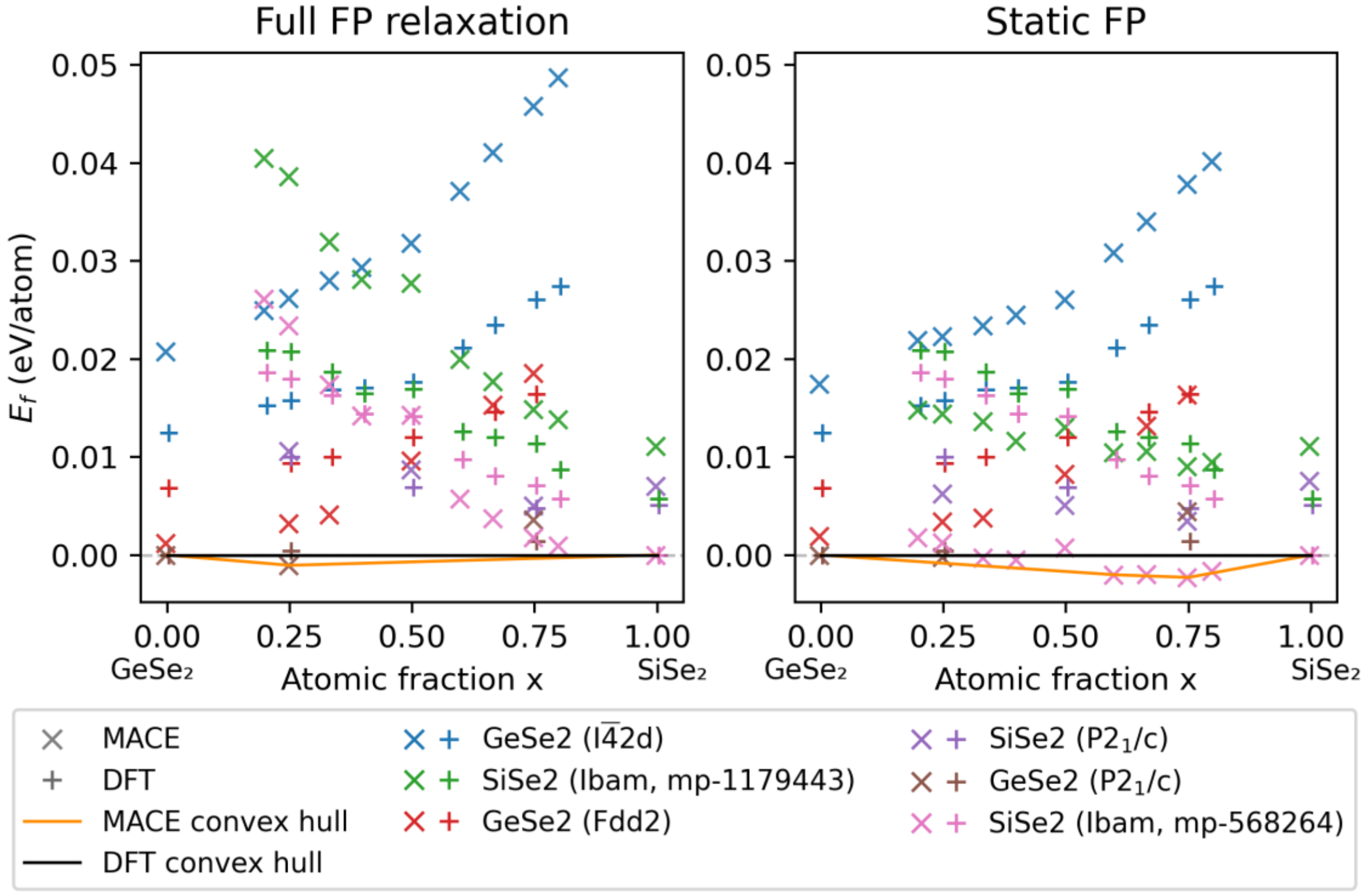

**Fig. 7: Convex-hull comparison between DFT and MACE for the $GeSe_2$–$SiSe_2$ tie-line system.** The convex hulls are shown for (left) full MACE relaxation, and (right) static MACE evaluations on DFT-relaxed structures (Methods). DFT (plus) and MACE (cross) data points at the same composition are slightly offset for visual clarity.

**Table 3:** Error-decomposition metrics for convex hull and relative phase stability. Implementation details for each metric are provided in the Methods.

| **Name** | **Metrics** |
|---|---|
| Average energy error | Energy error relative to DFT, reported as the MAE over all structures in the convex-hull subset. |
| Ground-state agreement: lowest-energy phase at each composition. | For each composition, the FP-predicted ground-state phase is compared with DFT. The reported metric is the fraction of compositions, across all tie-line systems, for which the two agree. |
| Within-phase hull-minimum agreement: lowest-energy composition within each phase. | For each phase, the FP-predicted minimum-energy composition is compared with DFT. The reported metric is the fraction of phases, across all tie-line systems, for which the two agree. |
| Hull-minimum agreement: global convex-hull minimum across the tie-line. | For each tie-line system, the FP-predicted convex-hull minimum is compared with DFT. The reported metric is the fraction of systems for which the two agree. |
| Structure relaxation error | For each structure, the RMSD between the FP-relaxed and DFT-relaxed structures is computed. The reported metric is the mean/maximum over all structures in the convex-hull subset. |

**Table 4:** Convex-hull accuracy and agreement metrics for FPs evaluated using full FP relaxation and static FP evaluations on DFT-relaxed structures (Table 3 and Methods).

| **FP** | **Full FP relaxation** | | | | **Static FP evaluations on the DFT-relaxed structures** | | | |
|---|---|---|---|---|---|---|---|---|
| | **Average energy error, MAE (meV/atom)** | **Ground-state agreement (%)** | **Within-phase hull-minimum agreement (%)** | **Hull-minimum agreement (%)** | **Average energy error, MAE (meV/atom)** | **Ground-state agreement (%)** | **Within-phase hull-minimum agreement (%)** | **Hull-minimum agreement (%)** |
| MACE | 53 | 68.3 | 62.0 | 40.9 | 15 | 70.2 | 73.6 | 40.9 |
| CHGNet | 242 | 53.4 | 40.5 | 18.2 | 246 | 55.9 | 48.8 | 22.7 |
| M3GNet | 49 | 55.3 | 47.1 | 22.7 | 33 | 55.9 | 59.5 | 22.7 |
| MatterSim | 58 | 59.6 | 59.5 | 27.3 | 22 | 62.1 | 71.1 | 27.3 |
| UMA | 83 | 62.7 | 60.3 | 36.4 | 43 | 74.5 | 78.5 | 40.9 |
| Orb | 81 | 59.6 | 58.7 | 27.3 | 46 | 71.4 | 76.0 | 31.8 |
| DPA4 | 84 | 66.5 | 60.3 | 45.5 | 45 | 77.6 | 78.5 | 54.5 |
| GRACE | 52 | 64.6 | 62.8 | 45.5 | 10 | 77.6 | 78.5 | 50.0 |
| M3GNet-MatPES | 117 | 50.3 | 55.4 | 22.7 | 109 | 49.1 | 57.0 | 22.7 |
| TensorNet-MatPES | 89 | 52.8 | 58.7 | 13.6 | 66 | 49.1 | 63.6 | 13.6 |
| MACE-MatPES | 89 | 49.7 | 59.5 | 18.2 | 55 | 53.4 | 71.9 | 18.2 |

We first evaluate the average energy error, which measures the overall agreement between FP and DFT energies, but does not directly indicate whether the correct phase stability is recovered. These errors are on the level of 49-242 meV/atom (Table 4) much larger than the typical energy MAEs of 18-35 meV/atom reported for some of these FPs on their respective in-distribution evaluation sets, such as MPTrj.[2,4,5] This shows that low average energy error on these in-distribution evaluation benchmarks does not translate to these complex substitutional and vacancy-ordered chalcogenide structures. Such errors are significant for phase-stability prediction because competing polymorphs and orderings in PCM systems often differ by only a few meV/atom.[36]

To test whether the FP identifies the correct most stable phase among all competing polymorphs, we evaluate the metric of ground-state agreement (Table 3 and Methods). For each composition along a tie line, the FP-predicted lowest-energy phase among the competing polymorphs is compared with the DFT ground-state phase. Across all tie-line systems, the metric reports the fraction of compositions for which the FP and DFT identify the same ground-state phase. Most models show moderate ground-state agreement ranging from approximately 50% to 68%. MACE gives the highest agreement at 68.3%, followed by DPA4, GRACE and UMA at 66.5%, 64.6% and 62.7%, respectively, while MACE-MatPES gives the lowest at 49.7% (Table 4).

For the series of substituted compositions within each phase along a tie line (Fig. 7), we next evaluate whether the FP correctly identifies the minimum-energy composition, quantified by the within-phase hull-minimum agreement metric (Table 3 and Methods). For each phase within a tie-line system (Fig. 7), the FP-predicted composition with the lowest energy is compared with the minimum-energy composition predicted by DFT. Across all tie-line systems in the dataset, we use the metric, within-phase hull-minimum agreement (Table 3 and Methods), to report the fraction of phases for which the two agree. Most FP models show agreement of approximately 55–63%, with GRACE performing best at 62.8%, followed by MACE at 62.0% and UMA and DPA4 at 60.3%. CHGNet shows the lowest agreement at 40.5% (Table 4).

Finally, by combining all competing phases and compositions, we evaluate the global convex-hull minimum using the hull-minimum agreement metric (Table 3 and Methods). This metric provides the most direct test of whether an FP preserves the relative thermodynamic stability required to reproduce the correct convex hull across an entire composition tie line. All models achieve less than 50% agreement, with DPA4 and GRACE performing best at 45.5%, followed by MACE at 40.9% and UMA at 36.4%, while the remaining models range from 13.6% to 27.3% (Table 4). Notably, M3GNet, GRACE, and MACE have comparable energy MAEs of 49, 52, and 53 meV/atom, yet achieve hull-minimum agreements of 22.7%, 45.5%, and 40.9%, respectively. Likewise, UMA, despite having a higher energy MAE, substantially outperforms M3GNet on this metric with 36.4% agreement. These comparisons demonstrate that average energy error alone is insufficient for evaluating FP performance in convex-hull construction and phase-stability prediction. Recovering the global convex-hull minimum remains a significant challenge for current FPs because it requires accurately reproducing the relative stabilities of all competing phases and compositions across highly substituted and defect-containing structures.[23,35]

### *2.2.2. Energy rankings of elemental orderings*

**Table 5:** Error-decomposition metrics for energy rankings of elemental orderings. This analysis was conducted on the ordering groups, each consisting of 20 lowest-energy DFT orderings for a fixed composition and phase (Methods).

| **Name** | **Metrics** |
|---|---|
| Average energy error | Energy error relative to DFT, reported as the MAE over all structures in the ordering subset. |
| Top-1 accuracy | Fraction of ordering groups for which the FP identifies the same lowest-energy configuration (LEC) as DFT.[23] |
| Recall@k | Fraction of the k lowest-energy DFT orderings retained among the top k FP-ranked orderings within each ordering group. The reported metric is the mean over all ordering groups. |
| Spearman's ρ | Rank correlation between FP and DFT orderings within each ordering group. The reported metric is the mean over all ordering groups. |
| Rate of ranking errors | Fraction of configuration pairs whose FP and DFT rankings disagree, within each ordering group. The reported metric is the mean over all ordering groups. |
| $\Delta E^{\mathrm{DFT}}$ of misranked pairs | DFT energy difference between configurations that the FP misranks.[23] For each ordering group, the mean and maximum $\Delta E^{\mathrm{DFT}}$ over its misranked pairs are computed and then averaged across all ordering groups. |

**Table 6:** Elemental-ordering ranking metrics for FPs evaluated using full FP relaxation and static FP evaluations on the DFT-relaxed structures. Ranking metrics are macro-averaged over ordering groups and the average energy error is pooled over all structures in the ordering subset (Table 5 and Methods). Arrows denote whether higher (↑) or lower (↓) values indicate better performance.

| protocols | FP | Average energy error, MAE (meV/atom) ↓ | Top-1 accuracy (%) ↑ | Recall@3 (%) ↑ | Recall @10 (%) ↑ | Spearman's ρ ↑ | Rate of ranking errors (%) ↓ | Mean/max $\Delta E^{\mathrm{DFT}}$ (meV/atom) ↓ |
|---|---|---|---|---|---|---|---|---|
| Full FP relaxation | MACE | 43 | 23.6 | 32.8 | 60.9 | 0.28 | 38.9 | 12 / 34 |
| | CHGNet | 272 | 8.2 | 20.1 | 52.8 | 0.09 | 46.8 | 13 / 39 |
| | M3GNet | 56 | 7.2 | 18.6 | 51.3 | 0.05 | 48.1 | 14 / 42 |
| | MatterSim | 47 | 18.0 | 31.1 | 58.7 | 0.24 | 41.0 | 12 / 36 |
| | UMA | 65 | 24.3 | 34.6 | 61.9 | 0.31 | 37.6 | 12 / 35 |
| | Orb | 65 | 25.2 | 33.7 | 61.3 | 0.29 | 38.5 | 12 / 35 |
| | DPA4 | 65 | 22.0 | 34.0 | 61.2 | 0.30 | 38.4 | 12 / 35 |
| | GRACE | 43 | 24.9 | 35.8 | 61.8 | 0.32 | 37.4 | 11 / 34 |
| | M3GNet-MatPES | 95 | 10.2 | 21.0 | 54.8 | 0.12 | 45.7 | 13 / 39 |
| | TensorNet-MatPES | 69 | 12.8 | 25.4 | 56.1 | 0.17 | 43.6 | 12 / 37 |
| | MACE-MatPES | 69 | 21.3 | 32.7 | 60.5 | 0.28 | 39.1 | 11 / 35 |
| Static FP evaluations on the DFT-relaxed structures | MACE | 14 | 46.6 | 54.5 | 72.7 | 0.56 | 26.6 | 5 / 15 |
| | CHGNet | 270 | 20.3 | 29.7 | 59.0 | 0.24 | 40.6 | 10 / 29 |
| | M3GNet | 41 | 22.3 | 28.7 | 58.7 | 0.23 | 41.0 | 10 / 30 |
| | MatterSim | 19 | 44.3 | 50.1 | 70.4 | 0.51 | 29.2 | 6 / 17 |
| | UMA | 35 | 55.7 | 65.4 | 79.9 | 0.69 | 19.7 | 4 / 10 |
| | Orb | 37 | 54.4 | 60.4 | 77.7 | 0.64 | 22.3 | 5 / 12 |
| | DPA4 | 36 | 57.0 | 64.0 | 79.9 | 0.69 | 19.7 | 4 / 10 |
| | GRACE | 10 | 58.7 | 65.7 | 80.1 | 0.70 | 19.2 | 4 / 10 |
| | M3GNet-MatPES | 94 | 19.7 | 29.8 | 60.0 | 0.26 | 40.2 | 10 / 30 |
| | TensorNet-MatPES | 52 | 27.9 | 41.4 | 65.3 | 0.38 | 34.8 | 8 / 23 |
| | MACE-MatPES | 43 | 41.0 | 48.0 | 70.0 | 0.50 | 29.6 | 6 / 18 |

Here we evaluate whether FPs can reproduce the energy ranking of different elemental orderings within the same phase and composition, which is required in practical studies of identifying the thermodynamically preferred ordering among many competing configurations. As shown by Liu et al.,[23] such ranking errors cannot be fully captured by average energy errors and must be evaluated using dedicated ranking metrics.

For this analysis, we use the ordering subset, consisting of 305 ordering groups, each containing 20 elemental or vacancy orderings of the same phase at a fixed composition, down selected from the dataset (Methods). For each ordering group, the FP-predicted energy ranking after full FP relaxation is compared with the corresponding ranking given by DFT energies. Then the ranking metrics (Table 5) are used to quantify whether the relative energy ranking among all evaluated elemental orderings is preserved.

We evaluate complementary ranking metrics (Table 5) that assess recovery of the lowest-energy configuration (LEC) (Top-1 accuracy) and retention of the k lowest-energy DFT orderings (Recall@k) as well as the correct prediction of the relative-energy ranking hierarchy (i.e. Spearman's ρ). Overall, the ordering performance is limited across all models. Top-1 accuracy is low even for the best-performing FPs, reaching only 25.2% for Orb and 24.9% for GRACE (Table 6). GRACE performs best across most ranking metrics, with the highest Recall@3 and Spearman's ρ and the lowest rate of ranking errors. UMA ranks second on most of these metrics and highest on Recall@10, despite having a higher energy MAE than MACE and M3GNet. Consistent with Section 2.2.1, low average energy error does not necessarily imply good relative-energy ranking.

A ranking error is defined as when the FP and DFT disagree on the energy ranking of a pair of configurations. Following Liu et al,[23] we quantify the rate of ranking errors as the fraction of misranked pairs among all pairs within an ordering group (Table 5 and Methods). All tested FP models show substantial ranking errors in this benchmark, with rates ranging from 37.4% for GRACE to 48.1% for M3GNet. These results indicate that even the current best model tested here misranks a large fraction of pairwise comparisons (Table 6). In addition to the rate of ranking errors, $\Delta E^{\mathrm{DFT}}$ measures the severity or magnitude of the misranking errors: misranking two nearly degenerate configurations is less consequential than misranking configurations separated by a large DFT energy difference. We report the mean and maximum $\Delta E^{\mathrm{DFT}}$ averaged over all ordering groups (Table 5 and Methods). The reported values are similar across all models.

*2.2.3. Structural relaxation effects*

The convex-hull and ordering analyses in Sections 2.2.1 and 2.2.2 are conducted for structures with full FP relaxation, in which FPs are used to relax doped, substituted, or defect-containing structures before evaluating their energies, consistent with common computational workflows.[9,17,36] The accuracy of the FP predictions for these cases therefore depends on two factors: (1) the intrinsic accuracy of the FP energies for a given atomistic configuration, and (2) differences in the optimized structures reached during FP relaxation, which can drive configurations to different local minima. To specifically test the first effect, namely intrinsic FP energy and ranking errors, we conduct static FP evaluations on the DFT-relaxed structures. The same convex-hull and ordering metrics are then evaluated using the resulting FP energies on the DFT structures (Tables 4 and 6).

**Table 7:** Structural agreement metrics between FP-relaxed and DFT-relaxed structures for the same dataset used in the convex-hull energy analysis. RMSD is computed between each FP-relaxed structure and the corresponding DFT-relaxed one, for structures that satisfy atom-to-atom mapping (Methods). Mean and maximum values across all mapped structures are reported, along with the fraction of structures below each RMSD cutoff.

| FP | Map success (%) | Mean/max RMSD (Å) | RMSD < 0.05 Å (%) | RMSD < 0.10 Å (%) | RMSD < 0.20 Å (%) |
|---|---|---|---|---|---|
| MACE | 89.9 | 0.28 / 1.92 | 24.0 | 44.5 | 65.5 |
| CHGNet | 87.3 | 0.37 / 2.28 | 11.9 | 26.5 | 49.5 |
| M3GNet | 88.8 | 0.37 / 2.02 | 10.2 | 21.1 | 39.4 |
| MatterSim | 89.9 | 0.26 / 2.09 | 22.0 | 44.5 | 67.6 |
| UMA | 89.9 | 0.29 / 2.69 | 24.4 | 45.4 | 66.1 |
| Orb | 89.4 | 0.30 / 2.63 | 20.8 | 40.4 | 63.3 |
| DPA4 | 89.6 | 0.27 / 2.56 | 24.7 | 45.8 | 65.6 |
| GRACE | 89.6 | 0.26 / 2.59 | 27.1 | 48.2 | 67.7 |
| M3GNet-MatPES | 89.8 | 0.36 / 2.41 | 12.3 | 26.3 | 51.3 |
| TensorNet-MatPES | 89.9 | 0.29 / 2.19 | 15.6 | 36.7 | 63.1 |
| MACE-MatPES | 89.6 | 0.27 / 2.31 | 23.6 | 44.5 | 64.5 |

When FP and DFT predictions are compared on the same structures, energy MAEs are significantly lower for most models, and ground-state and within-phase hull-minimum agreement also improve for several FP models (Table 4). Most ranking metrics, such as Top-1 accuracy and rank correlations, also improve for static evaluations, but remain far from perfect (Table 6). The rate of ranking errors remains substantial, and most FPs show little improvement in hull-minimum agreement. These results demonstrate that the errors identified in Sections 2.2.1 and 2.2.2 arise from both intrinsic energy-ranking errors and structural-relaxation errors.

To specifically test the effect of structural relaxation, we compare FP-relaxed and DFT-relaxed structures by measuring the displacement of each atom from its counterpart, reported as a per-structure RMSD for those where an atom-to-atom mapping can be established (Table 7 and Methods). Mapping success is approximately 87–90% across models, indicating that 10–13% of FP-relaxed structures differ significantly from the DFT minimum structures as a result of FP relaxation. Among the successfully mapped structures, the mean RMSD is 0.26–0.37 Å, with 39–68% falling within 0.20 Å of DFT and only 10–27% falling below 0.05 Å, showing that FP relaxation generally preserves the overall structural framework but still produces non-negligible atomic displacements relative to DFT. The maximum RMSD reaches 1.92–2.69 Å. These structural deviations reflect errors introduced during FP structural relaxation, likely arising from the force errors discussed in Section 2.1. The same relaxation error also underlies the endpoint-structure deviations in FP-NEB calculations (Section 2.3). Improving FP performance for substituted and defect-containing structures therefore requires not only more accurate relative energies but also more reliable forces to ensure that structural relaxation converges to the correct local minimum.

*2.2.4. Summary*

Together, these analyses decompose FP errors in thermodynamic-stability tasks into relative-energy ranking and structural-relaxation contributions and resolve the ranking errors across competing orderings, phases, and compositions. Models with similar average energy errors often exhibit substantially different performance in preserving the relative thermodynamic stability of competing phases, compositions, and elemental orderings. These decomposition error metrics identify whether task-level errors arise from incorrect relative energies among competing structures or from structural deviations introduced during FP relaxation, which cannot be inferred from average energy error alone.

Among the evaluated models, GRACE and MACE perform consistently well across the convex-hull and ordering benchmarks, while UMA achieves comparable ranking performance despite a higher average energy error. Models trained on the MatPES dataset show no consistent improvement over their non-MatPES counterparts, indicating that broader sampling of the potential-energy surface alone is insufficient to improve relative-energy ranking. This is likely because current FP training datasets generally do not specifically target highly substituted, vacancy-containing, and competing elemental orderings, making these benchmarks a challenging out-of-distribution evaluation for all current FPs. Improving FP performance for these applications will therefore require training datasets that include more representative competing phases, substitutional and vacancy orderings, and defect-containing structures.

### 2.3. Potential energy surfaces by Nudged Elastic Band (NEB) calculations

Nudged elastic band (NEB) calculations are widely used to evaluate atom-, ion-, and defect-diffusion pathways and energy barriers in solid materials.[37,38] However, DFT-based NEB is computationally expensive, especially when many materials or paths must be evaluated. FPs offer a potential alternative to DFT for accelerating these calculations.[8] Prior studies have reported systematic underestimation of predicted barriers for some FPs in NEB calculations.[19,24,25] Here, we apply error decomposition to the full FP-NEB workflow. Beyond evaluating migration-barrier accuracy, error-decomposition metrics (Table 8) resolve FP-NEB failures according to their source, including the underlying PES, endpoint relaxation, and along-path NEB optimization.

The test dataset comprises DFT-NEB calculations for 154 ion-migration pathways selected and computed from the ion-migration dataset of Saravanan et al. [39] (Methods). FP-NEB calculations are conducted with FPs as they would be used in practice, referred to as the full FP-NEB workflow (Methods): endpoint structures are first relaxed using the FP (with a force convergence threshold of 0.002 eV/Å), intermediate images are generated by linear interpolation, and the full NEB optimization is then performed using the FP (with a force convergence threshold of 0.05 eV/Å and a maximum of 1000 NEB steps). In addition to the full FP-NEB calculations, we also perform static FP evaluations on the DFT-NEB image structures to evaluate the FP error on the same DFT potential-energy surface and to isolate errors introduced by FP endpoint relaxation and NEB optimization (Section 2.3.4). The resulting FP energy profiles, migration barriers, and endpoint energies are compared directly with the corresponding DFT-NEB results.

**Table 8:** Error-decomposition metrics for PES by NEB calculations. Implementation details for each metric are provided in the Methods.

| **Name** | **Metrics** |
|---|---|
| Non-converged paths | Fraction of FP-NEB calculations that do not satisfy the NEB convergence criterion before reaching the maximum number of optimization steps. All subsequent error metrics are evaluated only for converged NEB paths. |
| Barrier error | The forward and backward barrier errors are computed relative to DFT for each path. The reported metric is the MAE/RMSE over all converged paths. |
| Endpoint-energy ranking agreement | Fraction of paths for which FP and DFT identify the same lower-energy endpoint between the two endpoints of the migration path, or both classify the two endpoints as equal in energy. |
| Endpoint-energy difference error | For each path, the error in the endpoint energy difference is computed relative to DFT. The reported metric is the MAE/RMSE over converged paths. |
| Energy-profile shape agreement | Fraction of paths for which the FP reproduces the Normal-Hill energy profile, as in DFT-NEBs. |
| Integrated energy-profile difference | For each path, the integrated absolute energy difference between FP and DFT-NEB energy profiles along the normalized reaction coordinate is computed. The reported metric is the MAE/RMSE over converged paths. |
| Endpoint-structure relaxation error | For each endpoint structure, the RMSD between FP-relaxed and DFT-relaxed endpoint structures is computed. The reported metric is the mean/maximum over all endpoint structures of converged paths. |
| Force errors on FP-NEB path | Mean force-magnitude error $\lvert\Delta\lvert F\rvert\rvert$ and force-angle error $\Delta\theta$ across all atoms for each image structure of the final FP-NEB path. |
| Force errors on DFT-NEB path | Mean force-magnitude error $\lvert\Delta\lvert F\rvert\rvert$ and force-angle error $\Delta\theta$ across all atoms for each image structure of the final DFT-NEB path. |

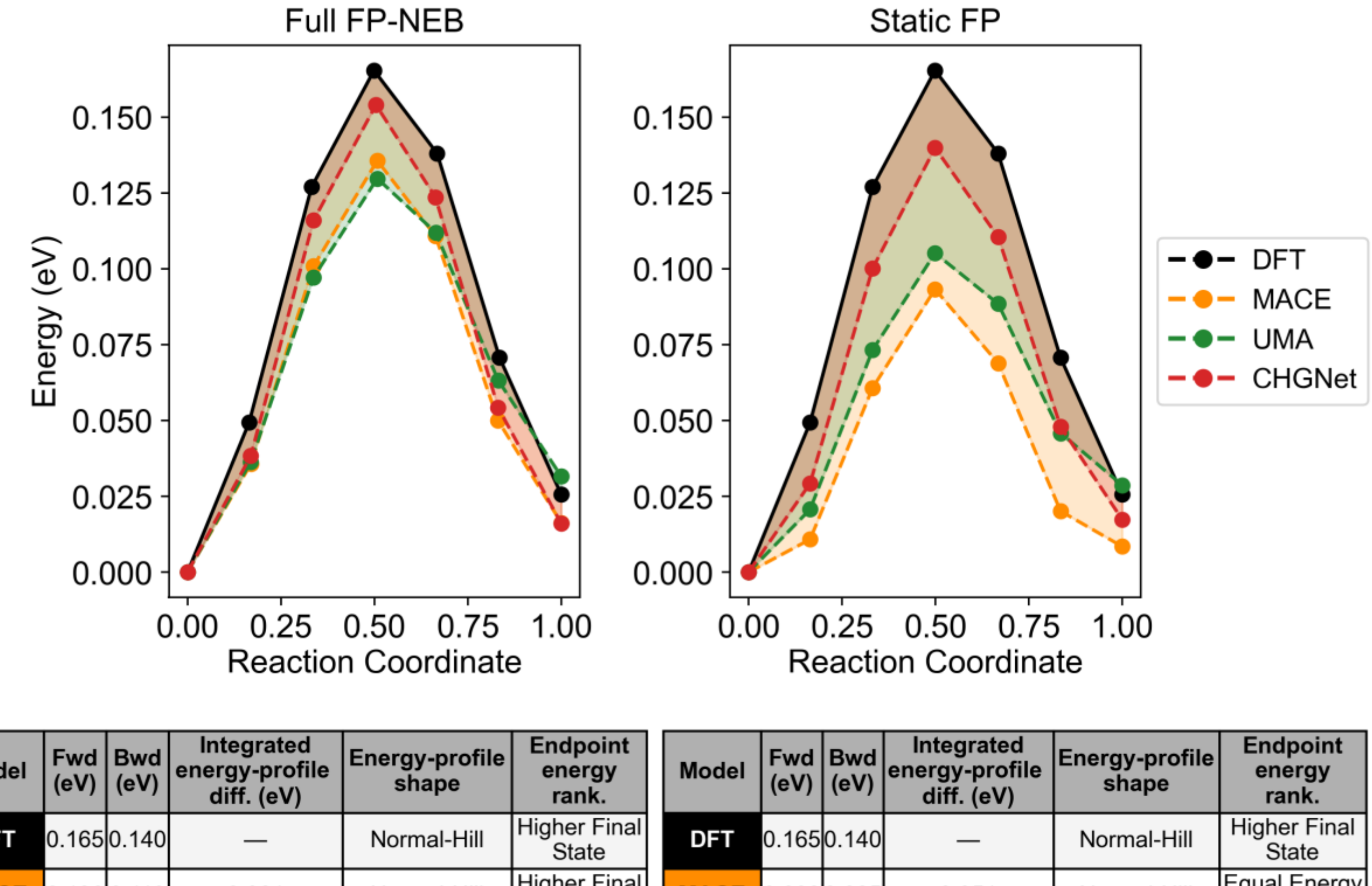

| Model | Fwd (eV) | Bwd (eV) | Integrated energy-profile diff. (eV) | Energy-profile shape | Endpoint energy rank. |
|---|---|---|---|---|---|
| DFT | 0.165 | 0.140 | — | Normal-Hill | Higher Final State |
| MACE | 0.136 | 0.119 | 0.021 | Normal-Hill | Higher Final State |
| UMA | 0.130 | 0.098 | 0.019 | Normal-Hill | Higher Final State |
| CHGNet | 0.154 | 0.138 | 0.013 | Normal-Hill | Higher Final State |

| Model | Fwd (eV) | Bwd (eV) | Integrated energy-profile diff. (eV) | Energy-profile shape | Endpoint energy rank. |
|---|---|---|---|---|---|
| DFT | 0.165 | 0.140 | — | Normal-Hill | Higher Final State |
| MACE | 0.093 | 0.085 | 0.051 | Normal-Hill | Equal Energy Endpoints |
| UMA | 0.105 | 0.076 | 0.036 | Normal-Hill | Higher Final State |
| CHGNet | 0.140 | 0.123 | 0.021 | Normal-Hill | Higher Final State |

**Fig. 8: Comparison of DFT and FP-predicted NEB energy profiles for the Oct–Oct migration pathway along the c-axis in $Li_3YCl_6$ (ICSD 29966).** The left panel shows full FP-NEB calculation results, and the right panel shows static FP evaluations on the DFT-NEB structures (Methods). The tables summarize the selected metrics for this path.

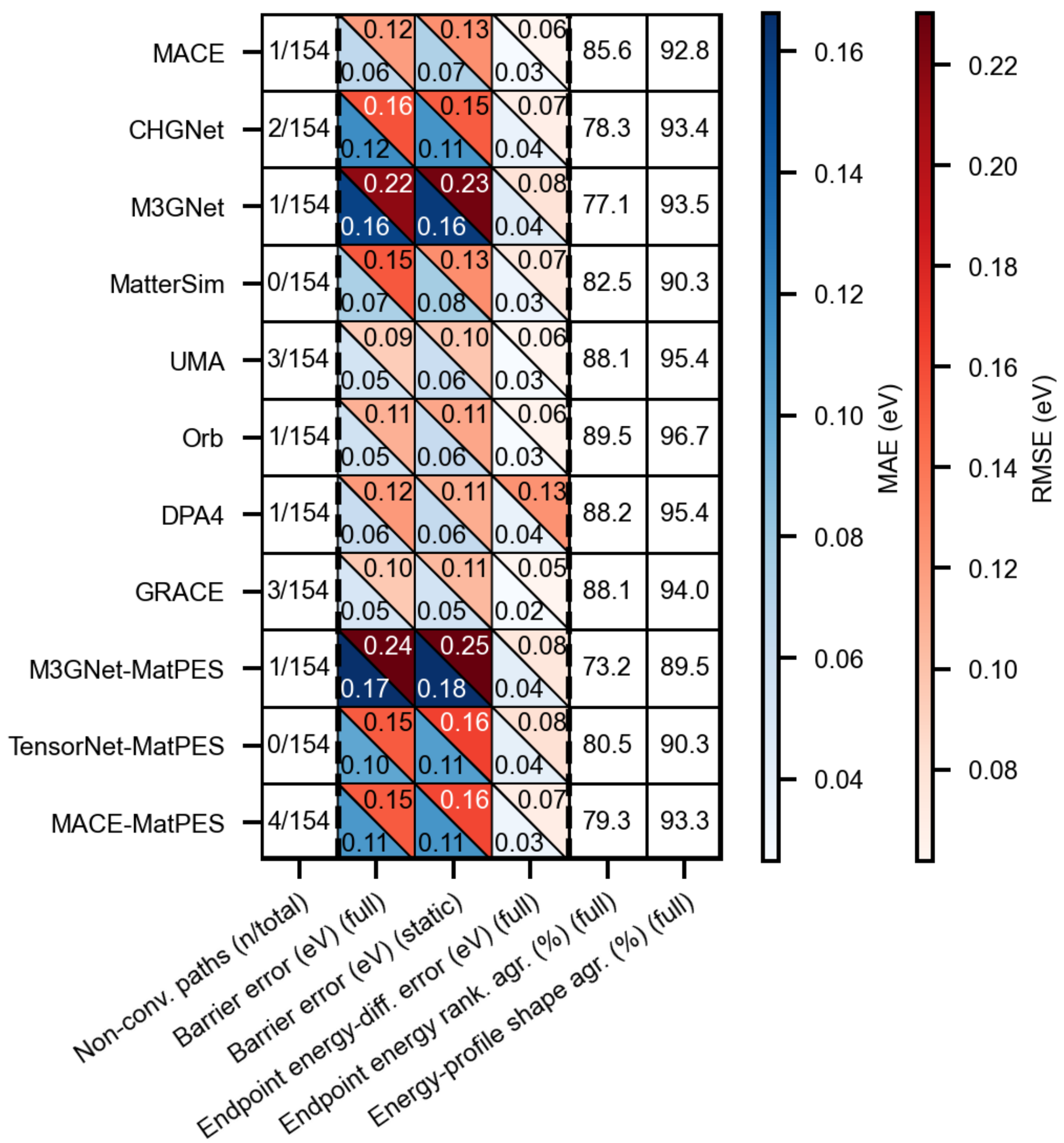

**Fig. 9: Summary of key metrics for all FP models on the NEB dataset.** MAE/RMSE values are shown in the lower/upper triangles, respectively. The same error metrics by static FP evaluations on the DFT-NEB paths are provided in Supplementary Fig. 16.

### *2.3.1. Failures in NEB calculations*

We first evaluate whether the full FP-NEB workflow produces converged and physically meaningful migration pathways. Across the 154 DFT-NEB pathways, a small but non-negligible number of full FP-NEB calculations, ranging from 0/154 to 4/154 across FP models, fail to reach the NEB force convergence criterion (Fig. 9). These non-converged paths are not reliable final NEB results and are therefore excluded from the barrier error analysis in the following sections (2.3.2-2.3.4).[40,41] Although the fraction of non-converged paths is low, it may indicate underlying deficiencies in force prediction or endpoint relaxation, and it provides a metric for assessing FP errors in practical NEB workflows. The possible origins of non-convergence, including endpoint relaxation differences and along-path force errors, are analyzed in Section 2.3.4.

In addition to numerical non-convergence, FP-NEB calculations can produce erroneous energy profiles. We classify an energy profile as “Normal-Hill” if at least one intermediate image has higher energy than both endpoints, as is typical for a migration energy profile. Converged FP-NEB paths that do not satisfy this criterion fall into two failure classes: (1) “Abnormal”, where no intermediate image exceeds both endpoint energies and at least one falls to or below the lower endpoint, corresponding to a negative barrier energy profile; and (2) “Invalid”, where an image adjacent to an endpoint has lower energy (Methods). Energy-profile shape agreement is defined as the fraction of converged FP-NEB paths that produce a Normal-Hill profile. All tested FPs achieve between 89.5% and 96.7% energy-profile shape agreement (Fig. 9), indicating that approximately 3–10% of paths do not exhibit a hill-like barrier profile. These cases represent qualitative NEB failures. Having identified non-converged and qualitatively abnormal paths, we next quantify barrier errors for the converged FP-NEB paths.

### *2.3.2. Barrier errors*

For converged paths, we quantify the accuracy of FP-predicted migration barriers from full FP-NEB calculations (Methods). Because the two endpoints can have different energies, each NEB path has both forward and backward barriers. We evaluate both forward and backward barrier errors, referenced to the lower- and higher-energy endpoints, respectively (Methods). For Abnormal profiles identified in Section 2.3.1, the barrier evaluation yields negative barriers, which produce higher barrier error and are therefore appropriately penalized in the MAE/RMSE values.

Barrier MAEs range from approximately 0.05 to 0.17 eV across models (Fig. 9). GRACE, UMA, and Orb achieve the lowest full FP-NEB barrier MAEs of approximately 0.05 eV, followed by DPA4, MACE, and MatterSim at 0.06–0.07 eV. TensorNet-MatPES,

MACE-MatPES, and CHGNet reach 0.10–0.12 eV, while M3GNet and M3GNet-MatPES show the largest errors at 0.16 and 0.17 eV. RMSE values are higher than MAE across all models, indicating the presence of some high-error FP-NEB results. Most models tend to underestimate barriers (Supplementary Figs. 17 and 18), consistent with the systematic underestimation of barriers reported in prior studies.[19,24,25] The integrated energy-profile difference, which integrates the absolute energy differences across all images along the reaction coordinate (Table 8 and Methods), shows similar trends of barrier errors (Supplementary Fig. 19).

*2.3.3. Endpoint energy ranking*

**Table 9:** Endpoint-structure relaxation error for converged FP-NEB paths. Mean and maximum RMSD are evaluated across all endpoint structures in the dataset, together with the fraction of endpoint structures below each RMSD cutoff.

| Potential | Mean/max RMSD (Å) | RMSD < 0.05 Å (%) | RMSD < 0.10 Å (%) | RMSD < 0.20 Å (%) |
|---|---|---|---|---|
| MACE | 0.06 / 0.73 | 72.5 | 86.6 | 94.1 |
| CHGNet | 0.08 / 0.66 | 54.6 | 78.3 | 91.4 |
| M3GNet | 0.13 / 0.66 | 37.9 | 61.4 | 78.1 |
| MatterSim | 0.07 / 0.76 | 64.0 | 83.4 | 91.2 |
| UMA | 0.05 / 0.62 | 76.8 | 87.4 | 96.7 |
| Orb | 0.05 / 0.60 | 78.4 | 89.5 | 96.1 |
| DPA4 | 0.06 / 0.76 | 76.8 | 86.9 | 95.4 |
| GRACE | 0.05 / 1.08 | 84.1 | 90.1 | 96.4 |
| M3GNet-MatPES | 0.16 / 0.83 | 27.9 | 50.5 | 73.8 |
| TensorNet-MatPES | 0.08 / 0.96 | 53.9 | 79.2 | 94.2 |
| MACE-MatPES | 0.06 / 0.83 | 68.0 | 85.7 | 94.7 |

The FP misprediction of the relative energies of the initial and final endpoints also leads to barrier errors (Fig. 8), because barriers are referenced to endpoint energies. Because ion migration typically involves hopping into a neighboring vacant site, endpoint energy errors are closely related to the errors in vacancy-ordering ranking discussed in Section 2.2.2. The endpoint energy-ranking agreement metric is evaluated to determine whether the FP preserves the relative stability of the two endpoints. This metric is defined as the fraction of paths for which the FP identifies the same lower-energy endpoint as DFT, or for which the FP and DFT both classify the two endpoints as equal in energy (Table 8 and Methods). This metric varies substantially across models, ranging from 73.2% for M3GNet-MatPES to 89.5% for Orb (Fig. 9), indicating that FPs misrank the relative endpoint energies for a substantial fraction of paths. In addition, we evaluate the endpoint energy-difference error, which measures how accurately the FP captures the energy difference between the two endpoints relative to DFT (Table 8 and Methods). For the tested FPs, endpoint energy-difference MAEs are approximately 0.02–0.04 eV, with RMSEs of 0.05–0.13 eV. These metrics characterize the contribution of the endpoint-energy errors to the barrier errors.

*2.3.4. Source of errors*

The errors discussed above can arise from various stages of the full FP-NEB workflow, which requires accurate FP relaxation of the endpoints and accurate forces throughout the subsequent NEB optimization. Here, we analyze three sources of error: (1) intrinsic inaccuracy in the FP description of the PES, revealed by static FP evaluations on the DFT-NEB images; (2) endpoint-structure relaxation error, where FP-relaxed endpoints drift away from their DFT-relaxed counterparts; and (3) force errors along the FP-NEB path, where incorrect forces on atoms of the image structures along the path can result in the wrong optimized path.

To separate intrinsic FP errors in the PES from errors introduced during the full FP-NEB workflow, such as endpoint relaxation or NEB path optimization, we perform static FP evaluations on the images of the final DFT-NEB path. FP energies and forces are computed on the final DFT-NEB images to directly assess how well FPs reproduce the DFT PES. Static barrier errors are generally similar to those from full FP-NEB calculations for most FP models (Fig. 9), indicating that intrinsic PES inaccuracy is a major source of barrier errors. We also evaluate the force errors on DFT-NEB path (Table 8 and Methods), and a path with a larger static barrier error can exhibit larger force errors (Supplementary Fig. 20), indicating inaccuracies in the PES gradient.

We next analyze the endpoint-structure relaxation, which is the first step of a full FP-NEB workflow and may significantly affect the subsequent NEB path optimization. We quantify this error using RMSD between FP-relaxed and DFT-relaxed endpoints across all converged paths (Table 9). Most FPs reproduce the endpoint structures reasonably well. UMA, Orb, and GRACE show the smallest mean endpoint RMSDs of 0.05 Å, with MACE, DPA4, and MACE-MatPES close behind at 0.06 Å, whereas M3GNet and M3GNet-MatPES show larger mean RMSDs of 0.13 and 0.16 Å, respectively. Because NEB endpoints typically represent single-vacancy or defect configurations, these relaxation errors are analogous to the structural deviations observed for vacancy-ordered structures in Section 2.2.3. The force errors responsible for this behavior are likely related to the low fraction of highly accurate force predictions discussed in Section 2.1.2.

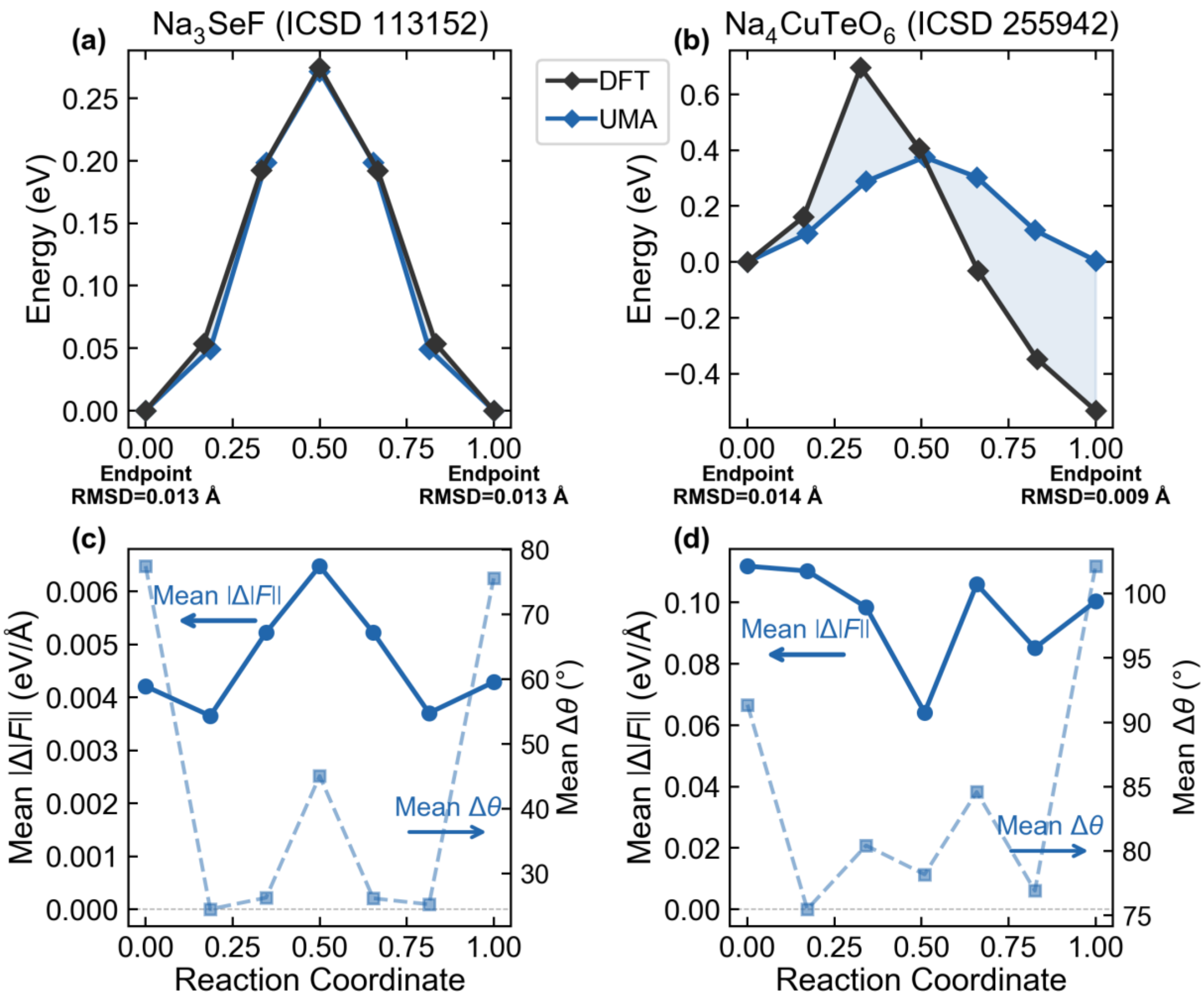

**Fig. 10: Along-path force-error on FP-NEB pathways.** (a, b) FP-NEB energy profiles for representative low- and high-barrier-error pathways. (c, d) Mean force-magnitude and force-angle errors over all atoms for each image. Large barrier errors can arise from along-path force errors even when endpoint structures are well reproduced (low RMSD).

We next evaluate the force errors on FP-NEB path (Table 8 and Methods), computed by performing DFT static calculations on each image structure of the final FP-NEB path and comparing FP and DFT forces on the same structures. These errors are related to the FE-atom force errors discussed in Section 2.1.4, because NEB image structures along the migration pathway are often far-from-equilibrium configurations. As shown by two representative paths (Fig. 10), the low-barrier-error path has small force errors along the entire path, whereas the high-barrier-error path exhibits substantially larger force errors at several image structures.

Together, the error-decomposition metrics (Table 8) show that full FP-NEB errors arise from multiple contributions, including intrinsic inaccuracies in the FP prediction of the PES, endpoint-relaxation errors, and along-path force errors during NEB optimization. Similar barrier errors can therefore arise from different underlying error sources that barrier MAE/RMSE alone cannot distinguish. Despite these PES and barrier errors, preliminary MD simulations of Li diffusion in LYC (Supplementary Fig. 23) show that several FPs still predict activation energies comparable to AIMD. However, substantial fluctuations and deviations from linear Arrhenius behavior remain. This suggests that agreement with AIMD activation energies alone should not be taken as evidence that an FP accurately reproduces the underlying PES or will perform reliably for other computational tasks.

## 3. Discussion and conclusion

### 3.1. From average error to application-oriented error-decomposition

Across the three computational tasks examined in this work, FP-DeErr demonstrates application-oriented error decomposition for force prediction, thermodynamic stability and relative-energy prediction, and NEB calculations. Average energy and force errors alone do not reliably predict FP performance for these tasks. Although models with low MAE or RMSE are often regarded as having reached near-DFT accuracy,[2,4,5] our results show that such models can still exhibit substantial errors in structural relaxation and MD simulations, phase-stability and elemental-ordering prediction, and NEB calculations. This limitation arises because different tasks are governed by different physical quantities, which are not directly captured by average errors. In some tasks, the atomistic configurations where these quantities are most consequential are also underrepresented in typical test sets, so their errors are diluted by the easier majority (the dilution effect). The three applications demonstrate complementary forms of error decomposition: force errors are resolved across prediction-accuracy regimes and physically consequential atom subsets; thermodynamic-stability errors are resolved across relative-energy comparisons and structural-relaxation contributions; and FP-NEB errors are resolved across the underlying PES and the stages of the computational workflow. Together, these applications establish a general strategy for application-oriented FP evaluation: identify the physically meaningful quantities governing task success, focus on the configurations or computational stages where errors in these quantities are most consequential, and devise diagnostic metrics that quantify these error components separately rather than averaging them over the full dataset or workflow.

***Force prediction***. Average force errors are low for nearly all current FPs, but these low values can be partly attributed to the dilution effect: because near-equilibrium, easy-to-predict atoms dominate the test dataset, the average error is diluted and can mask poor performance in the smaller, more consequential subset of atoms where accurate force prediction actually matters for downstream simulations. The error-decomposition metrics introduced in Section 2.1 (Table 1) fall into three categories that capture physically consequential cases. First, large-force-error atoms are the most damaging, because a single occurrence can destabilize an MD trajectory or misdirect a relaxation or NEB path. Both the fraction of such atoms and the magnitude of their errors should therefore be considered when evaluating practical FP reliability. Second, many FP computational tasks require highly accurate forces, well below the magnitude of errors that dominate the average. Such highly accurate predictions are relatively rare, and the average error does not reflect whether FP can achieve them, so a separate metric is needed. Third, far-from-equilibrium (FE) atoms are critical to many atomistic phenomena and are systematically underrepresented in near-equilibrium structure-relaxation datasets (e.g., MPTrj), although they are better represented in datasets such as MatPES and OMat24. Together, these results show that the force performance of current FPs is far from saturated for practical applications. A low average force error can arise largely from the dilution effect, and these three complementary metrics reveal key deficiencies that the average conceals, providing a more physically meaningful evaluation of force prediction for practical atomistic simulations using FPs.

***Phase stability and elemental ordering.*** Even when FPs report low average energy error across a dataset with diverse structures and compositions, our results show that an FP can still fail to correctly identify the relative thermodynamic stabilities among different phases, compositions, and elemental orderings. Even a modest FP energy error can erroneously invert the predicted ranking of two competing phases, orderings, or defect configurations, leading to incorrect predictions in practical computational tasks such as defect energetics, substitutional and vacancy ordering, Monte Carlo simulations, and thermodynamic stability for materials screening.[23,35] The thermodynamic-stability error-decomposition metrics introduced in Section 2.2 resolve relative-energy errors across competing orderings, phases, compositions, and separate these from the effects of structural relaxation. Ordering-ranking metrics evaluate whether an FP correctly ranks many competing elemental and defect/vacancy orderings at fixed composition and phase. Ground-state agreement and within-phase hull-minimum agreement evaluate whether an FP correctly identifies the most stable phase at a given composition and the minimum-energy composition within each phase, respectively. Hull-minimum agreement is a challenging test of convex-hull construction, requiring the FP to correctly identify the true global minimum across all competing phases and compositions. Our results show that preserving this relative thermodynamic stability remains a common weakness among current FPs, likely reflecting insufficient representation of such configurations, including diverse vacancy orderings and concentrations, in current training datasets. Therefore, for FP tasks that require reliable phase-stability prediction, convex-hull construction, or elemental-ordering studies, these application-oriented metrics should be explicitly evaluated.

***Ion migration by NEB calculations.*** Beyond barrier errors,[19,24,25] our results reveal additional failure modes: NEB paths can fail to converge or produce qualitatively incorrect energy profiles. The metrics introduced in Section 2.3 capture these failures through the fraction of non-converged paths, energy-profile shape agreement, and endpoint energy-ranking agreement, while endpoint-structure relaxation error and along-path force errors trace these failures back to specific stages of the FP-NEB workflow.

These errors connect directly to the FP errors in force prediction and structural relaxation identified in Sections 2.1 and 2.2. Notably, despite these barrier- and path-level errors, FP-based MD simulations of ionic diffusion still yielded activation energies broadly comparable to AIMD simulations, suggesting that averaged transport properties can be less sensitive to errors of individual ion migration paths or local PES. These metrics provide the diagnostic detail needed to guide future FP improvement for reliable migration-barrier prediction.

The error-decomposition framework of FP-DeErr is not specific to the three tasks considered here. For a new computational application or task, the same framework can be used to identify its governing physical quantities, focus error evaluation on consequential or error-prone configurations or workflow stages, and quantify these errors using corresponding diagnostic metrics.

### 3.2. Implications for future foundation potential development

Optimizing only conventional MAE or RMSE is unlikely to produce consistent improvements across diverse computational tasks, because different tasks depend on different physically meaningful quantities that average errors do not directly measure. Application-oriented error decomposition provides specific targets for FP development by identifying which physical quantities are inaccurate and the configurations where these errors arise and become consequential. Rather than optimizing only aggregate MAE/RMSE or task-level scores, FP development can target these deficiencies through improvements in training data, model architecture, or training objectives.

Evaluation datasets should be developed together with the corresponding error-decomposition metrics, as shown in the various analyses of FP-DeErr. Datasets should contain a representative fraction of the physically consequential and error-prone configurations for the corresponding computational task, while the associated metrics explicitly evaluate FP performance on those configurations. The datasets used in our work were based on or generated from realistic computational studies to reflect practical FP applications,[10,11,36] but they represent only a subset of the materials classes and configurations encountered in these tasks, and are therefore not yet sufficient for truly foundational benchmarking. As FPs continue to improve, benchmark datasets will need to evolve toward broader materials systems, a wider range of application scenarios, and increasingly challenging but physically relevant configurations. Such datasets can also provide targeted training data for the deficiencies identified through error decomposition.

To facilitate broad adoption of our error-decomposition framework, we share the complete code for both running the FP calculations and computing all error-decomposition metrics.[42] The code is designed so that new FP models, metrics, and datasets can be incorporated easily. Our FP-DeErr framework will be extended as new FPs are developed and additional application-oriented error-decomposition metrics and more rigorous evaluation datasets are introduced by the community. We also introduce a public FP-DeErr leaderboard[43] to provide standardized comparison of FP performance and encourage community contributions as the framework extends. Beyond the application-oriented error-decomposition metrics developed for the three representative computational tasks in this work, there remains a need for similar metrics across many other practical applications of FPs.[6,18,31,32] Extending the FP-DeErr framework to these computational tasks is therefore an important direction for future work.

In summary, we recommend that application-oriented error-decomposition metrics be reported alongside conventional energy and force MAE/RMSE, so that FP evaluation captures not only overall prediction accuracy and task-level performance but also the physically consequential failure modes that determine practical reliability. By identifying what errors arise and where they become consequential, application-oriented error decomposition provides actionable targets for improving FP training, evaluation, and model development. The application-oriented error-decomposition framework, diagnostic metrics, evaluation datasets, and extensible infrastructure introduced in FP-DeErr will guide future FP evaluation and development toward improving practical reliability across computational tasks.

## 4. Methods

### 4.1. Foundation Potentials (FPs)

**Table 10:** Specifications of FP Model used in this work.

| FP | Model & Version | Model Size | Training dataset | Training dataset size |
|---|---|---|---|---|
| MACE[44] | >=v0.3.10 (mace-mpa-0-medium) | 9.06M | MPTrj + sAlex | ~3.5M |
| CHGNet[4] | v0.3.0 | 412.5K | MPTrj | ~1.58M |
| M3GNet[2] | MP-2021.2.8-PES | 288.2K | MP-2021.2.8 | ~176.6K |
| MatterSim[13] | MatterSim-v1.0.0-5M | 4.55M | MatterSim dataset, composition not disclosed | ~6M[45] |
| UMA[12] | s-1p1 | 146.5M | OC20 + ODAC23 + OMat24 + OMC25 + OMol25 | ~500M |
| Orb[14] | orb-v3-conservative-inf-omat-20250404 | 25.5M | OMat24 (AIMD subset) | ~55M |
| DPA4[15] | DPA4-Plus-OMat24-v20260805 | 8.85M | OMat24 | ~100.6M[46] |
| GRACE[16] | GRACE-3L-OMAT-large-ft-AM | 42.1M† | fine-tuned using sAlex + MPTrj * | ~12.0M (fine-tuning only)[47] |
| M3GNet-MatPES[11] | v2025.1 | 664.2K | MatPES-PBE | ~435K |
| TensorNet-MatPES[11] | v2025.1 | 837.9K | MatPES-PBE | ~435K |
| MACE-MatPES[44] | >=v0.3.10 (MACE-matpes-pbe-omat-ft) | 9.06M | fine-tuned using MatPES-PBE * | ~435K |
| M3GNet-MatPES[11] (r2SCAN) | v2025.1 | 664.2K | MatPES-r2SCAN | ~388K |
| TensorNet-MatPES[11] (r2SCAN) | v2025.1 | 837.9K | MatPES-r2SCAN | ~388K |
| MACE-MatPES[44] (r2SCAN) | >=v0.3.10 (MACE-matpes-r2scan-omat-ft) | 9.06M | fine-tuned using MatPES-r2SCAN * | ~388K |

* pre-trained using OMat24

† For GRACE, the parameter count excludes the optional uncertainty estimator, which was disabled in our calculations.

Here, we systematically evaluated eleven FP models (Table 10), including three trained on MatPES dataset. Because the MatPES-trained models are evaluated on their training distribution (Table 11), we also test all models on the OMat24 rattled-1000 dataset as an out-of-distribution benchmark for some models (Supplementary Table 2).

**Table 11:** Evaluation matrix of FPs across benchmarks. A check mark indicates that the model was evaluated on that dataset. OOD = out-of-distribution, indicating the test set falls outside the model's training distribution. The MatPES models trained on r2SCAN dataset is also evaluated on the MatPES-r2SCAN dataset.

| FP | MatPES-PBE | OMat24 rattled-1000 | Convex Hull / ordering / NEB |
|---|---|---|---|
| MACE[44] | ✓ (OOD) | ✓ (OOD) | ✓ (OOD) |
| CHGNet[4] | ✓ (OOD) | ✓ (OOD) | ✓ (OOD) |
| M3GNet[2] | ✓ (OOD) | ✓ (OOD) | ✓ (OOD) |
| MatterSim[13] | ✓†(OOD) | ✓†(OOD) | ✓ (OOD) |
| UMA[12] | ✓ (OOD) | ✓ (training) | ✓ (OOD) |
| Orb[14] | ✓ (OOD) | ✓ (OOD) | ✓ (OOD) |
| DPA4[15] | ✓ (OOD) | ✓ (training) | ✓ (OOD) |
| GRACE[16] | ✓ (OOD) | ✓ (pretraining)* | ✓ (OOD) |
| M3GNet-MatPES[11] | ✓ (training) | ✓ (OOD) | ✓ (OOD) |
| TensorNet-MatPES[11] | ✓ (training) | ✓ (OOD) | ✓ (OOD) |
| MACE-MatPES[44] | ✓ (training) | ✓ (pretraining) * | ✓ (OOD) |

* pre-trained using OMat24. MACE–MatPES was subsequently fine-tuned using MatPES-PBE, whereas GRACE was fine-tuned using sAlex and MPTrj.
† MatterSim’s training data is undisclosed.

### 4.2. Datasets

**MatPES dataset:** MatPES dataset is used to evaluate the force predictions in Section 2.1. MatPES dataset is a collection of representative equilibrium and off-equilibrium structures selected from finite-temperature MD trajectories and augmented with Materials Project ground-state structures, with energies, forces, and stresses labeled by DFT static calculations[11]. The MatPES dataset includes a substantial fraction of far-from-equilibrium (FE) atoms (26.8 %). While in the main text, the models and dataset based on PBE[48] functional is used, the same evaluation based on MatPES-r2SCAN dataset and their trained models can be found in Supplementary Table 1.

**OMat24 rattled-1000 dataset:** The force prediction evaluation in Section 2.1 also uses the rattled-1000 validation subset of the OMat24 dataset[10] (~117K structures), which includes a set of non-equilibrium structures generated by applying random atomic displacements and cell deformations and selecting from a Boltzmann-like distribution at a sampling temperature of 1000 K, labeled with static DFT-PBE calculations. This subset provides an additional test of FP force predictions in addition to the MatPES-PBE dataset, as some models are evaluated on their own training dataset (Table 11).

**Chalcogenide PCM dataset:** For the phase stability and elemental ordering in Section 2.2, we use two subsets of data derived from the ternary chalcogenide PCM dataset of Adams et al.[36] Ternary structures are generated by interpolating compositions along binary–binary chalcogenide tie lines (a total of 22) in the composition space of Si, Ge, Sn, Sb, Bi, Ga, In and Ti with Se or Te, using isovalent cation or anion substitutions or aliovalent cation substitutions with charge compensation by cation or anion vacancies. For each phase and composition, the symmetrically distinct elemental and vacancy orderings are enumerated, and all structures are relaxed with DFT-PBE calculations. For testing FP convex-hull and phase-stability predictions in Section 2.2.1, we used the convex-hull subset with only the lowest-energy ordering of each phase and composition, giving 597 unique structures across 22 tie-line systems. This subset comprises 561 ternary structures and 36 binary endpoint structures. For the elemental-ordering ranking in Section 2.2.2, the dataset is organized into ordering groups, each consisting of elemental or vacancy orderings of the same phase at a fixed composition. This ordering subset retains only groups with at least 20 orderings and selects the 20 lowest-energy orderings per group, giving 6,100 structures across 305 ordering groups.

**NEB dataset:** For testing FP-NEB calculations in Section 2.3, we use a subset of 154 Li- and Na-ion migration pathways from the ion-migration dataset of Saravanan et al.,[39] derived from a high-throughput screening study of fast Li- and Na-ion conductors. These 154 pathways span 106 structurally distinct materials and a range of chemistries, including oxides (32.1%), sulfides (20.8%), halides, and multi-anion compounds. The DFT-NEB reference calculations for these pathways were performed in this work (see DFT-NEB calculations).

### 4.3. DFT calculations

All DFT calculations in this work were performed with the Vienna Ab initio Simulation Package (VASP)[49] using the projector augmented-wave (PAW)[50] approach and the PBE[48] generalized gradient approximation, with parameters largely following the Materials Project input set.[51] An energy cutoff of 520 eV, Gaussian smearing with a width of 0.05 eV, and spin polarization were used. Hubbard U corrections were applied where applicable using the Materials Project U values and the Dudarev scheme.[52] Γ-centered k-point meshes were generated with pymatgen.[53] The convergence parameters used in all static DFT calculations were set to be consistent with the Materials Project.

### 4.4. DFT-NEB calculations

For each pathway, in the NEB dataset the two endpoint structures were first relaxed at fixed cell shape and volume with convergence criteria of $1\times10^{-6}$ eV (electronic) and $1\times10^{-4}$ eV (ionic) and a maximum of 1000 ionic steps. Five images were then interpolated between the two endpoints, giving seven images in total, and the NEB optimization used a force convergence criterion of 0.05 eV/Å and a maximum of 1000 NEB steps.

### 4.5. Force evaluation metrics

For each atom in each structure of the datasets, FP forces were computed by static FP evaluations on the structures using ASE.[54] Evaluated on each atom, the force error was decomposed into force-magnitude error,

$$\Delta|F| = |F_{\mathrm{FP}}| - |F_{\mathrm{DFT}}| \tag{1}$$

and force-angle error,

$$\Delta\theta = \cos^{-1}\frac{F_{\mathrm{FP}}\cdot F_{\mathrm{DFT}}}{|F_{\mathrm{FP}}||F_{\mathrm{DFT}}|} \tag{2}$$

defined as the angle between the FP and DFT force vectors (Fig. 2a). We also computed the force-vector error,

$$e_{\mathrm{vec}} = \|F_{\mathrm{FP}} - F_{\mathrm{DFT}}\|, \tag{3}$$

defined as the norm of the difference between the FP and DFT force vectors. Atoms with zero FP or DFT force were excluded from all metrics, since the force angle is undefined in these cases. All metrics in Section 2.1 (except for the all-atom MAE/RMSE values in Fig. 5) were evaluated only for atoms with $|F_{\mathrm{DFT}}| > 0.01$ eV/Å, because for near-zero DFT forces, small absolute differences in the force components can produce large changes in the force angle.

*Average error.* The average error was evaluated using $\Delta|F|$ and $\Delta\theta$, reported as MAE/RMSE across all atoms or a selected subset. The full error distributions were characterized by the cumulative distribution functions (CDFs) of $|\Delta|F||$, $\Delta\theta$, and $e_{\mathrm{vec}}$.

*Highly accurate force predictions.* We computed the fraction of atoms with small $|\Delta|F||$, and small $\Delta\theta$, using different thresholds (e.g. $|\Delta|F||$ from 0.01 to 0.5 eV/Å); $|\Delta|F||$< 0.01 eV/Å was used as the criterion for highly accurate force-magnitude predictions. Joint force magnitude-angle accuracy was evaluated as the fraction of atoms simultaneously satisfying $|\Delta|F||$ thresholds (0.01 to 0.5 eV/Å) and force-angle error thresholds of $\Delta\theta$ < 1° or $\Delta\theta$ < 20°.

*Large-force-error atoms.* The atoms with large $|\Delta|F||$ and $\Delta\theta$, using $|\Delta|F||$ thresholds (>0.5 eV/Å to 10 eV/Å) are classified as large-force-error atoms.

Far-from-equilibrium (FE) atoms. Far-from-equilibrium (FE) atoms were defined as atoms with $|F_{\mathrm{DFT}}| > 1$ eV/Å. For FE atoms, we reported $\Delta|F|$ and $\Delta\theta$ MAE/RMSE and evaluated the relative force-magnitude error,

$$r_F = \frac{|\Delta|F||}{|F_{\mathrm{DFT}}|} \tag{4}$$

to assess relative force accuracy in the FE regime.

**4.6. Convex-hull and elemental-ordering metrics**

For the full FP relaxation workflow, atomic positions were relaxed from their initial geometries at fixed cell shape and volume using the FP with the FIRE optimizer[55] in ASE,[54] with a force convergence threshold of 0.01 eV/Å and a maximum of 10,000 steps. The initial geometries are the unrelaxed structures, except for the 36 binary endpoint structures, which are taken from Materials Project entries. The resulting FP-relaxed structures and energies were compared with the corresponding DFT-relaxed references. Static FP evaluations are also conducted and compared (4.6.3).

*4.6.1. Convex hull and relative phase stability*

The MAE of the per-atom energy difference between FP and DFT was pooled over the 597 unique structures of the convex-hull subset defined in Section 4.2. Ground-state agreement compares per-atom energies of all phases at a fixed composition to determine whether FP and DFT identify the same lowest-energy phase at each composition. Compositions represented by only one phase were excluded.

For the hull-minimum metrics, which compare minima across different compositions along a tie line, $E_{\mathrm{f}}(x)$ is the formation energy per atom relative to the two binary endpoints,

$$E_{\mathrm{f}}(x) = E(x) - [(1-x)E_0 + xE_1] \tag{5}$$

where $E(x)$ is the per-atom energy at atomic fraction $x$ along the tie-line, and $E_0$ and $E_1$ are the per-atom energies of the endpoint structures at $x$ = 0 and $x$ = 1, respectively. The convex hull from FP or DFT was normalized to its own endpoint energies.

Within-phase hull-minimum agreement measures whether FP and DFT identify the same minimum-energy composition within each phase, and the hull-minimum agreement measures whether they identify the same global convex-hull minimum across all phases in a tie-line system. When neither FP nor DFT identifies an intermediate composition as the hull minimum, they are counted as agreeing if they identify the same most stable phase at both binary endpoints.

*4.6.2. Energy rankings of elemental orderings*

FP and DFT energy rankings were compared within each ordering group defined in Section 4.2. Following Liu et al.,[23] the rate of ranking errors was computed as

$$f_{err} = \frac{2\, n_d}{n(n-1)}, \tag{6}$$

where *n* is the number of structures in the ordering group and $n_d$ is the number of misranked configuration pairs, with a pair of configurations $(i, j)$ counted as misranked when $\left(E_i^{\mathrm{DFT}} - E_j^{\mathrm{DFT}}\right)\left(E_i^{\mathrm{FP}} - E_j^{\mathrm{FP}}\right) \leq 0$. Whereas Liu et al. reported the rate per phase, we compute $f_{err}$ for each ordering group and report the average across all groups. For each misranked pair (i, j), we also computed the per-atom DFT energy difference,

$$\Delta E^{\mathrm{DFT}} = \left|E_i^{\mathrm{DFT}} - E_j^{\mathrm{DFT}}\right|, \tag{7}$$

where $E^{\mathrm{DFT}}$ is the per-atom DFT energy for each configuration. For each ordering group, the mean and maximum $\Delta E^{\mathrm{DFT}}$ over misranked pairs were computed and then averaged across all ordering groups.

Additional ranking metrics were computed from the sorted FP and DFT energy rankings. Top-1 accuracy is the fraction of ordering groups for which the FP correctly identifies the same lowest-energy configuration (LEC) as DFT. Recall@k measures the fraction of the k lowest-energy DFT orderings retained among the top k FP-ranked orderings. Spearman's ρ[56] measures the rank correlation between FP and DFT orderings and was computed with the built-in function in *SciPy*.[57] Recall@k and Spearman's ρ were computed within each ordering group and averaged over all groups.

*4.6.3. Structural relaxation effects*

Static FP evaluations on the DFT-relaxed structures were conducted, and the FP energies were compared directly with the corresponding DFT energies. Together with the full FP-relaxation results, this comparison separates intrinsic FP energy-ranking errors from errors introduced by FP structural relaxation. The structural agreement between FP-relaxed and DFT-relaxed structures was evaluated using RMSD quantified by the StructureMatcher function in pymatgen,[53] with a fractional length tolerance (ltol) of 0.5, a site tolerance (stol) of 0.5, an angle tolerance (angle_tol) of 10°, and primitive-cell reduction was disabled (primitive_cell=False). These looser tolerances than the pymatgen[53] defaults allow moderately distorted FP-relaxed structures to be mapped to their DFT counterparts, which determines mapping success. RMSD was computed only for mapped structures. The RMSD returned by StructureMatcher is dimensionless, computed after both cells are isotropically rescaled to a common volume so that structures are compared on a common length scale. We recovered the RMSD in physical units (Å) by multiplying by

$$\left(\frac{V_{\mathrm{geom}}}{N}\right)^{\frac{1}{3}}, \tag{8}$$

where N is the number of atoms and

$$V_{\mathrm{geom}} = \sqrt{V_{\mathrm{FP}} V_{\mathrm{DFT}}} \tag{9}$$

is the geometric mean of the FP- and DFT-relaxed volumes.

### 4.7. NEB evaluation metrics

The full FP-NEB workflow was conducted using MatCalc.[58] For each pathway, the two endpoint structures were relaxed using the FP with the MatCalc RelaxCalc interface and the BFGS optimizer in ASE,[54] with a force convergence threshold of 0.002 eV/Å, a maximum of 500 steps, and fixed cell shape and volume. Five intermediate images were then generated by linear interpolation using pymatgen,[53] consistent with the DFT-NEB reference paths. The climbing-image FP-NEB optimization was performed using MatCalc NEBCalc interface with the BFGS optimizer in ASE,[54] with a force convergence threshold of 0.05 eV/Å and a maximum of 1000 steps. FP-NEB calculations that did not reach this convergence criterion after reaching the maximum steps were counted as non-converged paths, and reported as a fraction of the 154 pathways. Converged NEB paths were further analyzed for errors.

#### *4.7.1. Failures in NEB calculations*

Energy profiles were classified as Normal-Hill, Abnormal, or Invalid as follows. We first classify a profile as Abnormal when no intermediate image rose above both endpoints and at least one fell to or below the lower-energy endpoint, giving negative or zero forward and backward migration barriers. A profile was classified as Invalid if an image adjacent to an endpoint had lower energy than that endpoint, indicating possible endpoint-relaxation error. All remaining profiles have at least one intermediate image above both endpoints and no endpoint-adjacent image below its endpoint, giving positive forward and backward migration barriers, and were classified as Normal-Hill. Energy-profile shape agreement was reported as the fraction of converged FP-NEB paths that reproduced the Normal-Hill profile of the DFT-NEB reference.

#### *4.7.2. Barrier errors*

For each converged path, $E_{\mathrm{initial}}$ and $E_{\mathrm{final}}$ denote the initial and final endpoint energies. The energy of the extremum image, $E_{\mathrm{ex}}$, was identified in one of two ways depending on the profile shape. When at least one intermediate image had energy above both endpoints, $E_{\mathrm{ex}}$ was set to the maximum intermediate-image energy, giving a positive migration barrier. When no intermediate image rose above both endpoints, $E_{\mathrm{ex}}$ was instead taken as the intermediate-image energy with the largest absolute deviation from either endpoint, which can give negative or zero barrier. The forward barrier was referenced to the lower-energy endpoint,

$$\text{Forward barrier} = E_{\text{ex}} - \min(E_{\text{initial}}, E_{\text{final}}) \tag{10}$$

and the backward barrier to the higher-energy endpoint,

$$\text{Backward barrier} = E_{\text{ex}} - \max(E_{\text{initial}}, E_{\text{final}}). \tag{11}$$

For each converged path, regardless of profile classification, the forward and backward barrier errors were computed relative to the corresponding DFT barriers, and MAE/RMSE were reported over the pooled set of forward and backward barrier errors.

The integrated energy-profile difference was evaluated by trapezoidal integration as

$$\int_0^1 |E_{\text{FP}}(\text{s}) - E_{\text{DFT}}(\text{s})| \, \text{ds}. \tag{12}$$

The reaction coordinate $s$ was scaled to the interval [0,1] following the standard convention as implemented in ASE. This quantity is the mean absolute energy difference along the path, in eV, and was reported as MAE/RMSE over converged paths.

*4.7.3. Endpoint energy ranking*

For each converged NEB path, the endpoint energy difference was defined as

$$\Delta E_{\text{endpoints}} = E_{\text{final}} - E_{\text{initial}}. \tag{13}$$

According to the sign of $\Delta E_{\text{endpoints}}$, each path was classified as higher initial endpoint, higher final endpoint, or equal endpoints (defined as $|\Delta E_{\text{endpoints}}| \leq 0.01$ eV). Endpoint energy-ranking agreement was the fraction of converged paths for which FP and DFT are classified the same. The endpoint energy-difference error was computed as

$$\Delta E_{\text{endpoints}}^{\text{FP}} - \Delta E_{\text{endpoints}}^{\text{DFT}} \tag{14}$$

and reported as MAE/RMSE over converged paths.

*4.7.4. Source of errors*

Endpoint-structure relaxation error was evaluated from the RMSD between each FP-relaxed endpoint structure and its corresponding DFT-relaxed endpoint structure, using the RMSD procedure described in Section 4.6.3. All endpoint structures were successfully mapped, except one for M3GNet-MatPES. We report the mean and maximum RMSD over all endpoint structures of all converged paths.

For the static FP evaluations on the DFT-NEB image structures, only static FP evaluations were performed on the DFT image structure with no FP endpoint relaxation or FP-NEB optimization. The energy-profile and barrier metrics and the FP forces were computed using the same definitions above. For force-error diagnostic, DFT static calculations were performed for each image structure, including the two endpoint structures of the selected final FP-NEB paths (see DFT calculations). For each image structure, the FP and DFT forces were compared, and the mean force-magnitude error $\left|\Delta|F|\right|$ and force-angle error $\Delta\theta$ were computed.

**Author contributions.** K.A. and Y.M. designed the study. K.A. developed FP-DeErr and performed the benchmarking calculations and analyses. R.S.S. generated the NEB dataset used in this work, and F.A. generated the ternary chalcogenide phase-change-material dataset. All authors contributed to the discussions. Y.M. supervised the project. K.A. and Y.M. wrote the manuscript.

**Code availability.** The complete FP-DeErr benchmarking code, including routines for running the FP calculations and evaluating all metrics, is publicly available through the FP-DeErr GitHub repository at https://github.com/mogroupumd/FP-DeErr.[42] A public FP-DeErr leaderboard is available at https://mogroupumd.github.io/FP-DeErr/.[43]

**Data availability.** The datasets used in this work are available from their original sources as specified in Methods. The DFT-NEB reference for the 154 ion-migration pathways evaluated in Section 2.3 is available at https://doi.org/10.6084/m9.figshare.33332610.[59] The convex-hull subset and ordering subset used in Section 2.2 are available at https://doi.org/10.6084/m9.figshare.33334329.[60]

**Competing interests.** The authors declare no competing interests.

**Acknowledgements.** The authors acknowledge the funding support from National Science Foundation Award# 2118838, 2329087, 2520978 and the computational facilities from the University of Maryland supercomputing resources.